\documentclass{emulateapj}		

\journalinfo{Tentatively scheduled for the March 2004 issue of The Astronomical Journal}	

\submitted{AJ in press}			

\shorttitle{Faint 6.7\,$\mu$m Galaxies and Stellar Mass Density}
\shortauthors{Y.~Sato et al.}

\begin{document}

\title{
FAINT \lowercase{6.7\,$\mu$m} GALAXIES AND THEIR CONTRIBUTIONS TO	
THE STELLAR MASS DENSITY IN THE UNIVERSE\footnotemark[1]}		
\footnotetext[1]{							
Based on observations with ISO,
an ESA project with instruments funded by ESA Member States
(especially the PI countries: France, Germany, the Netherlands
and the United Kingdom) and with the participation of ISAS and NASA.
}

\author{
Yasunori~Sato\altaffilmark{1,2},
Lennox L.~Cowie\altaffilmark{3},
Kimiaki~Kawara\altaffilmark{2},
Hideo~Matsuhara\altaffilmark{4},
Haruyuki~Okuda\altaffilmark{5},
David B.~Sanders\altaffilmark{3,8},
Yoshiaki~Sofue\altaffilmark{2},
Yoshiaki~Taniguchi\altaffilmark{6},
and
Ken-ichi~Wakamatsu\altaffilmark{7}
}

\altaffiltext{1}{
JSPS Research Fellow
}
\altaffiltext{2}{
Institute of Astronomy, University of Tokyo,
2-21-1 Osawa, Mitaka, Tokyo, 181-0015 Japan;
ysato@ioa.s.u-tokyo.ac.jp
}
\altaffiltext{3}{
Institute for Astronomy, University of Hawaii,
2680 Woodlawn Drive, Honolulu, HI 96822
}
\altaffiltext{4}{
Institute of Space and Astronautical Science (ISAS),
Japan Aerospace Exploration Agency,
3-1-1 Yoshinodai, Sagamihara, Kanagawa, 229-8510 Japan
}
\altaffiltext{5}{
Gunma Astronomical Observatory,
6860-86 Nakayama, Takayama, Agatsuma, Gunma, 377-0702 Japan
}
\altaffiltext{6}{
Astronomical Institute, Graduate School of Science, Tohoku University,
Aramaki, Aoba, Sendai, 980-8578 Japan
}
\altaffiltext{7}{
Faculty of Engineering, Gifu University,
1-1 Yanagido, Gifu, 501-1193 Japan
}
\altaffiltext{8}{
Max-Planck-Institut f\"{u}r extraterrestrische Physik,
D-85740, Garching, Germany
}

\begin{abstract}

We discuss the nature of faint 6.7\,$\mu$m galaxies
detected with the mid-infrared camera ISOCAM
on board the Infrared Space Observatory (ISO).
The 23\,hour integration on the Hawaii Deep Field SSA13
has provided a sample of 65 sources down to 6\,$\mu$Jy at 6.7\,$\mu$m.
For 57 sources, optical or near-infrared counterparts
were found with a statistical method.
All four \textit{Chandra} sources, three SCUBA sources,
and one VLA/FIRST source in this field
were detected at 6.7\,$\mu$m with high significance.
Using their optical to mid-infrared colors,
we divided the 6.7\,$\mu$m sample into three categories:
low redshift galaxies with past histories of rapid star formation,
high redshift ancestors of these,
and other star forming galaxies.
Rapidly star forming systems at high redshifts
dominate the faintest end.
Spectroscopically calibrated photometric redshifts
were derived
from fits to a limited set of template SEDs.
They
show a high redshift tail in their distribution
with faint ($<30$\,$\mu$Jy) galaxies at $z>1$.
The 6.7\,$\mu$m galaxies
tend to have brighter $K$ magnitudes and redder $I-K$ colors
than the blue dwarf population at intermediate redshifts.
Stellar masses of the 6.7\,$\mu$m galaxies were estimated from
their rest-frame near-infrared luminosities.
Massive galaxies ($M_\mathrm{star} \sim 10^{11}$\,M$_\odot$)
were found in the redshift range of $z=0.2$--3.
Epoch dependent stellar mass functions
indicate a decline of massive galaxies'
comoving space densities with redshift.
Even with such a decrease,
the contributions of the 6.7\,$\mu$m galaxies
to the stellar mass density in the universe
are found to be comparable
to those expected from UV bright galaxies
detected in deep optical surveys.

\end{abstract}


\section{INTRODUCTION}

The evolution of galaxies can be traced
in their spectral energy distributions (SEDs)
as a result of the aging and accumulation of stellar populations.
Because more massive stars evolve faster,
ultraviolet (UV) emission
becomes relatively weaker as time elapses,
provided that no further star formation occurs.
This sensitivity to on-going star formation
can be used to trace the star formation history in galaxies;
however, the UV light suffers from dust extinction.
Thus, optical observations of the redshifted UV emission from distant galaxies
need undesirable corrections
if we are to deduce the evolution of their star forming activity.

However, if we observe galaxies at a longer wavelength,
this undesirable situation is improved.
At rest near-infrared wavelengths,
much of the emission originates from low mass stars.
Their lifetime is comparable to the age of the universe;
thus, the effect of aging is much milder than that in the UV.
The complicating effects of dust extinction
are almost negligible at this wavelength.
Most of the changes in the near-infrared SEDs are caused
by the accumulation of stellar populations in galaxies.
These facts assure good accuracy in estimating
stellar masses of galaxies from their near-infrared luminosities.

The stellar mass
is one of the most fundamental quantities of galaxies.
This motivates the execution of galaxy surveys in the near-infrared.
For distant galaxies,
very sensitive surveys should be performed at a longer wavelength
in order to detect their rest-frame near-infrared light.
The mid-infrared camera ISOCAM \citep{CAA+96}
on board the Infrared Space Observatory \citep[ISO,][]{KSA+96}
has achieved such requirements for the first time.

Most of the deep ISOCAM surveys
have utilized two broad band filters:
LW2 (5--8.5\,$\mu$m) at 6.7\,$\mu$m
and LW3 (12--18\,$\mu$m) at 15\,$\mu$m
\citep{SEO+97,TCS+97,FHD+99,FHT+99,AMK+99,OMC+02}.
In particular, those at 15\,$\mu$m attracted much interest
\citep{GC00,FAC+01,FBR+03},
mainly due to the discovery of
a strongly evolving population of galaxies below 1\,mJy
\citep{ECF+99}.
The excess in number could be explained by
a large number of star forming galaxies at $z \sim 1$.

The 6.7\,$\mu$m cosmological observations with ISOCAM
resulted in generally fewer detections than at 15\,$\mu$m 
and thus attracted less attention.
This is because such observations should sample
a wavelength valley in galaxy SEDs between stellar and hot dust emission.
The passband of the LW2 filter matches
the location of the unidentified infrared band (UIB) emission
in the local universe.
Such dust emission shifts away from the passband with redshift
\citep{ACES99}.
At high redshifts $z>1$,
the major contribution to the LW2 band
becomes stellar emission in galaxies.
However, such emission from distant galaxies
requires extremely high sensitivities.
Based on the importance of detecting near-infrared stellar emission
from high redshift galaxies for estimating their stellar masses,
we conducted a very deep 6.7\,$\mu$m survey
in the Hawaii Deep Field SSA13
\citep{SKC+03}.
One sigma sensitivity reached 3\,$\mu$Jy,
which is smaller than the published values of 7\,$\mu$Jy
in \citet{TCS+97} and in \citet{ACES99}.
Most recently, \citet{MKM+03} reach
a depth similar to ours using a massive cluster lens.
In this paper, we discuss insights deduced from
the faint 6.7\,$\mu$m galaxies detected in the SSA13 field.

After describing the 6.7\,$\mu$m sample in Sect.~\ref{sect:samples},
we discuss the identification of multi-wavelength counterparts
(Sect.~\ref{sect:id}).
In Sect.~\ref{sect:nature},
the nature of the identified galaxies is examined
with the GRASIL star/dust SED model
\citep{SGBD98}.
Their stellar masses are obtained
using the rest-frame near-infrared luminosities,
and then we discuss the evolution of the stellar mass function
and the stellar mass density in the universe
(Sect.~\ref{sect:SM}).
Finally, we present a discussion (Sect.~\ref{sect:dis})
and conclusions (Sect.~\ref{sect:con}).
Throughout this paper,
we assume a flat universe with $\Omega_\mathrm{M}=0.3$, $\Omega_\Lambda=0.7$,
and $H_0=65\,\mathrm{km}\,\mathrm{s}^{-1}\,\mathrm{Mpc}^{-1}$.
All optical and near-infrared magnitudes in this paper are in the Vega system.

\section{THE \lowercase{6.7\,$\mu$m} SAMPLE}	
\label{sect:samples}

A deep mid-infrared survey has been conducted with the ISOCAM array
on board the Infrared Space Observatory (ISO).
The broadband filter LW2 (5--8.5\,$\mu$m)
with the reference wavelength of 6.7\,$\mu$m
was used to image a high galactic latitude region
in the Hawaii Deep Field SSA13.
Many raster observations totaling an observing time of 23\,hours
were combined to give a map with a nominal areal coverage of 16 arcmin$^2$
with the beam FWHM of 7.2 arcsec.
Details of the observations are given in \citet{SKC+03}.

With a non-uniform noise distribution over the map,
source detections were performed using a signal-to-noise ratio (S/N) map.
The noise map was created
from the standard deviations of the co-added raster images at each pixel.
Detection parameters were determined
by comparing numbers of detected sources at the positive and negative sides
of the map ($N_\mathrm{pos}$ and $N_\mathrm{neg}$, respectively)
with the same detection parameter set.
We have chosen the parameter set giving
the largest $N_\mathrm{pos}-N_\mathrm{neg}$,
where $N_\mathrm{pos}=65$ and $N_\mathrm{neg}=12$.
Total 6.7\,$\mu$m fluxes of the positive sources
resulted in a range from 6\,$\mu$Jy to 170\,$\mu$Jy
(Table~\ref{tab:id}).

By setting a threshold that can exclude all the negative sources,
we extracted a subsample of the positive sources.
This subsample, the primary sample,
consists of 33 sources having total fluxes larger than 12\,$\mu$Jy
and detection S/Ns larger than 4.3
(Table~\ref{tab:id}).
Because there should be no effects from spurious detections
in the primary sample,
our main results will be deduced based on the primary sample.
No corrections for the contamination of fake sources
are necessary for the primary sample
to derive integrated quantities
such as stellar mass functions and stellar mass densities
discussed in Sect.~\ref{sect:SM}.
In fact, galaxy number counts shown in \citet{SKC+03}
were obtained only with the primary sample.

The remaining 32 positive sources are put into the supplementary sample.
For the number of negative sources ($N_\mathrm{neg}=12$)
having comparably low significance values,
the number of true sources in the supplementary sample could be 20.
However, we expect more sources in the supplementary sample are real.
This is because the adopted image processing could produce
negative ghosts around bright sources,
and at least three of the negative sources were identified as such
\citep{SKC+03}.
Genuine sources in the supplementary sample
should provide meaningful information on the 6.7\,$\mu$m sources,
especially at the faintest flux levels.
Some of them will be mentioned in the following sections
with a caveat that there will be some effects from spurious sources.

In Table~\ref{tab:id}, sources in both samples are listed.
Here we also show the 12 negative sources as the negative sample.
All are to be examined with the source identification procedure
in the next section.

\section{SOURCE IDENTIFICATION}
\label{sect:id}

Counterparts of the 6.7\,$\mu$m sources were searched
at multiple wavelengths, from the X-ray to the radio.
We used two identification methods,
the probability ratio method in the optical and near-infrared, and the nearest neighbor search
at X-ray, submillimeter, and radio wavelengths.
The results are summarized in two tables,
Tables~\ref{tab:id} and \ref{tab:xsr}, respectively.

\subsection{Optical and Near-Infrared Identifications}
\label{sect:optid}

\subsubsection{The identification procedure}

Optical and near-infrared counterparts of the 6.7\,$\mu$m sources
were sought in the $K$, $I$, and $B$ bands.
These three band data were taken from the ground
\citep{CSHC96},
but we utilized a $HST$/WFPC2 catalogue in the $I_{814}$ band as well
\citep{CHS95}.
The catalogue limits were estimated as
$K=19.9$, $I_{814}=24.7$, $I=23.7$, and $B=25.5$ (3\,$\sigma$).
Photometric uncertainties were derived by summing relative and
absolute uncertainties quadratically.
Absolute photometric uncertainties were
estimated to be 0.1 magnitude for corrected aperture magnitudes
\citep{CGH+94}.
We do not distinguish between $I_{814}$ and $I$ magnitudes in the following,
because these magnitudes for the same sources in our sample
do not show significant differences.
At these magnitude limits,
surface densities become higher than that of the 6.7\,$\mu$m sources.
We then introduced the following two probabilities
to evaluate whether a candidate counterpart of magnitude $m_1$
at distance $r_1$ should be considered as a true association
\citep{MOS+97,FHD+99}.

Even for a true association between sources at two wavelengths,
there is expected to be a certain amount of displacement in their positions.
If we can neglect differences in light profiles at the two wavelengths,
a major cause for the displacement
will be measurement errors at both wavelengths.
Based on smaller beam sizes in the optical and near-infrared
($\sim 1$\,arcsec or less),
we only took account of measurement errors in the 6.7\,$\mu$m coordinates.
With an assumption that the errors follow a Gaussian distribution,
the probability for a true association $P_\mathrm{t}$ is defined as
\begin{equation}
P_\mathrm{t} = 1 - \sqrt{\frac{2}{\pi}} \int_{0}^{\alpha} \exp (- \frac{x^2}{2}) dx,
\end{equation}
where $\alpha(=r_1/r_0)$
is the displacement of a potential counterpart
normalized to the one sigma positional accuracy $r_0$
of the 6.7\,$\mu$m source in question.
This $r_0$ value was determined with Monte-Carlo simulations
in \citet{SKC+03}.

As long as the displacement has a finite value $r_1$,
we can not exclude the possibility
that an irrelevant source will be found at a distance smaller than $r_1$.
Such a chance event should follow Poisson statistics
for an area of ${\pi}{r_1}^2$.
Then the probability for a chance association $P_\mathrm{c}$ is derived as
\begin{equation}
P_\mathrm{c} = 1 - \exp (- {\pi}{r_1}^2 N(<{m_1})),
\end{equation}
where $N(<{m_1})$ is the surface density
of optical or near-infrared sources with magnitudes brighter
than $m_1$, the magnitude of a potential counterpart.
Because only two stars are expected in this 6.7\,$\mu$m sample
\citep{SKC+03},
$N(<m)$ was derived by integrating differential galaxy counts
shown in \citet{MSC+01} and in \citet{MIT+01}.

These two probabilities, $P_\mathrm{t}$ and $P_\mathrm{c}$,
were calculated for all potential counterparts
up to a distance of $3\,r_0$ from each 6.7\,$\mu$m position.
The large search radius was necessary to take account of
the large uncertainty in the determination of $r_0$
\citep{SKC+03}.
We then selected an optical or near-infrared counterpart
having the highest ratio of $P_\mathrm{t} / P_\mathrm{c}$.
If there is no candidate source with $P_\mathrm{t}/P_\mathrm{c}>1$,
the 6.7\,$\mu$m source was regarded as unidentified.
Assuming that SEDs of the 6.7\,$\mu$m sources are smoothly connected,
we performed the identification procedure starting from the nearest wavelength.
The adopted $K$ and $I_{814}$ band catalogues
do not cover the full 6.7\,$\mu$m field.
Taking also account of the deeper limit for the $I_{814}$ band catalogue,
the identification scheme was then set
from the $K$, $I_{814}$, $I$, to $B$ band.
If the identification at a particular band resulted in a success,
counterparts at the shorter wavelengths were found
with a nearest neighbor search.
A search up to 1.4 arcsec for it turned out to be enough.

\subsubsection{Identification results}
\label{idres}

\begin{figure*}
  \plotone{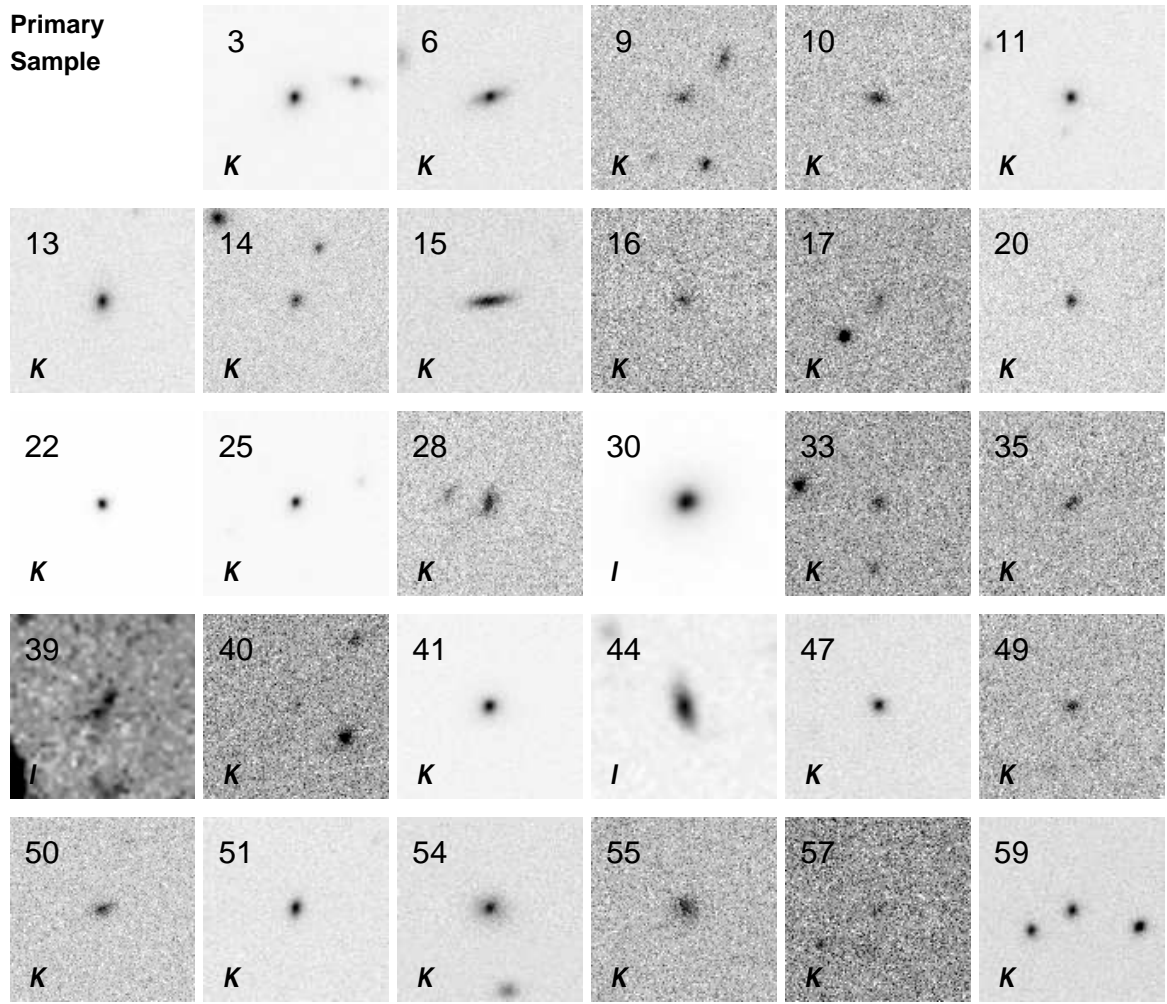}	
  \caption{
Thumbnail images of the identified 6.7\,$\mu$m sources.
The images were taken from
the data where the actual identifications were performed
(IDs in Table\ref{tab:id}).
This imaging band and the name of the 6.7\,$\mu$m source
were overlaid on each 18\,$\times$\,18\,arcsec panel,
an area of $3 \times 3$ pixels
for the ISOCAM 6 arcsec pixel-field-of-view mode.
The identified sources were centered in the images.
North is up and east to the left.
  }
\label{fig:id}
\end{figure*}

Of 65 positive sources, 57 were identified in the optical or near-infrared.
Fig.~\ref{fig:id} shows their appearances
at the wavelength where the probability calculations were performed.
In Table~\ref{tab:id},
identification band (ID), distance ($r_1$),
two probabilities ($P_\mathrm{t}$ and $P_\mathrm{c}$), and their ratio
are listed.
The coordinates of the identified sources have been updated with
those of the optical or near-infrared counterparts.
We also show the second highest ratio of $P_\mathrm{t} / P_\mathrm{c}$
($P_\mathrm{t}/P_\mathrm{c}^{(2)}$)
if other candidates were found in the search radius.
There are relatively few cases of ambiguous identifications
(i.e., where the second largest $P_\mathrm{t} / P_\mathrm{c}$
is of the same order of magnitude as the largest).
In particular, in the primary sample,
only one source (\#59) seems to fall in this category.
In the following, the identification results
for the primary and supplementary samples are examined separately.

For the primary sample,
we found that many sources were identified in the $K$ band.
All the sources that were identified at shorter wavelengths
have large $P_\mathrm{t}/P_\mathrm{c}$ values.
As indicated in Table~\ref{tab:prop},
they either lack $K$ data (\#30, \#39, and \#62)
or have blue colors (\#44 and \#64).
\citet{SCK+02} identified the source \#40
with a submillimeter source with a hard X-ray counterpart.
This identified source has
a smaller significance value ($P_\mathrm{t}/P_\mathrm{c} \sim 0.6$)
than the nominal threshold ($P_\mathrm{t}/P_\mathrm{c}=1$)
at the $K$ band.
We regard it as likely that the asymmetric profile of this 6.7\,$\mu$m source
\citep{SCK+02}
affected the probability calculation.
The only unidentified source \#27, having a high S/N=10,
lacks $K$ data
(Table~\ref{tab:prop})
and has a very red mid-infrared to optical color.

The lower significance of the supplementary sample
leads to poorer positional accuracy ($r_0 \sim$ 2--3 arcsec),
which made the identification more difficult.
Even in such circumstances,
78\,\% (25/32) of the supplementary sample were identified.
Because the supplementary and negative samples have the same S/N level,
the identification significance for the supplementary sample
can be checked by a comparison with the negative sample.
No source in the negative sample was identified in the $K$ band.
Thus the identifications of the supplementary sample
in the $K$ band are expected to be robust.
Three sources in the negative sample were identified
in the $I$ or $I_{814}$ band.
However,
most of the $I$ or $I_{814}$ band identifications in the supplementary sample
have better statistics,
i.e, larger $P_\mathrm{t}/P_\mathrm{c}$ values
than those for the $I$ or $I_{814}$ band identifications in the negative sample.
Here $I$ and $I_{814}$ counterparts should be treated separately
because of the difference in the catalogue depth.
The deeper $I_{814}$ sources
generally have smaller $P_\mathrm{t}/P_\mathrm{c}$ values
due to the larger surface density of sources.
The $I_{814}$ counterparts for sources \#21 and \#24
have larger $P_\mathrm{t}/P_\mathrm{c}$ values
than that of the $I_{814}$ counterpart for the negative source N7.
The $I$ counterparts for sources \#12, \#29, \#34, \#46, \#58, and \#60
have larger $P_\mathrm{t}/P_\mathrm{c}$ values
than that of the $I$ counterpart for the negative source N9,
which has a $P_\mathrm{t}/P_\mathrm{c}$ value larger than N10.
The remaining $I$ or $I_{814}$ band identifications
are for sources \#2, \#31, \#43, and \#52,
though most of them have slightly smaller $P_\mathrm{t}/P_\mathrm{c}$ values
than that of their corresponding negative source: N7 or N9.
Thus, we expect the number of erroneous identifications in the supplementary sample
will be quite small, four at most.
The number of unidentified sources in the supplementary sample (7)
is comparable to that in the negative sample (9).
This indicates that
some of the unidentified sources in the supplementary sample
could be spurious,
especially those detected in noisy regions of the map
(e.g. sources \#7 and \#37).
However, some could be very red sources like the source \#27 in the primary sample.

\citet{FHD+99} give $r_1$ and $P_\mathrm{c}$ values
for the identification of their 6.7\,$\mu$m sources.
Their values are generally larger than ours;
the median $r_1$ is 4.2 arcsec and $P_\mathrm{c}$ exceeds 0.3
in some cases.
We think that this may be partly due to their neglect of 
distortion corrections in data processing, which could be as large as one 
6 arcsec pixel in size.
They adopted 1.5 arcsec map pixels and used only 3--5 sources to determine
shifts between images.
The median FWHM of 6.7\,$\mu$m sources on their final map
was 11 arcsec, while ours was 7.2 arcsec.
We took account of distortion corrections
and used 0.6 arcsec map pixels
and 19 reference sources for the image registration
\citep{SKC+03}.
The difference mentioned above might not explain
all of the difference in their identification results.
Rather,
even the sources identified with the least significance in \citet{FHD+99}
have 6.7\,$\mu$m fluxes larger than ours
and correspondingly optical magnitudes brighter than ours.

\subsection{X-ray, Submillimeter, and Radio Identifications}

At the flux levels of the adopted X-ray, submillimeter, and radio catalogues,
the surface densities are lower than that of our mid-infrared sample.
Thus, source identification at these wavelengths
can be achieved using a simple method, a nearest neighbor search.

With the deep \textit{Chandra} observations of SSA13
\citep{MCBA00},
four X-ray sources were detected within the area of the ISOCAM survey.
Three of them were detected in the hard X-ray band (2--10\,keV)
with fluxes larger than
$4 \times 10^{-15}\mathrm{erg}\,\mathrm{s}^{-1}\,\mathrm{cm}^{-2}$,
while two of them were detected in the soft X-ray band (0.5--2\,keV)
with fluxes larger than
$8 \times 10^{-16}\mathrm{erg}\,\mathrm{s}^{-1}\,\mathrm{cm}^{-2}$
(Table~\ref{tab:xsr}).
Within 1.1\,arcsec of these four X-ray sources,
comparable to the positional accuracy of the \textit{Chandra} satellite,
we found 6.7\,$\mu$m counterparts \#14, \#17, \#40 and \#62.
Here we have used the 6.7\,$\mu$m coordinates
updated with the optical and near-infrared identifications above.

The Submillimetre Common-User Bolometer Array (SCUBA)
on the 15\,m James Clerk Maxwell Telescope (JCMT)
detected three 850\,$\mu$m sources down to 2\,mJy
in the deep SCUBA survey of SSA13
\citep{BCS99}.
Taking account of the very broad beam at 850\,$\mu$m (15\,arcsec FWHM),
all the three submillimeter sources were identified with
6.7\,$\mu$m sources \#28, \#40, and \#57
at distances of up to 7.7\,arcsec
\citep{SCK+02}.
Note that one of the submillimeter sources (\#40)
was also detected in the hard X-ray.

The VLA/FIRST survey at 1.4\,GHz detected one 3\,mJy source
in the 6.7\,$\mu$m SSA13 map
\citep{WBHG97}.
This source is also listed in the VLA/NVSS catalogue
\citep{CCG+98}.
The 6.7\,$\mu$m counterpart is assigned to the source \#20
at a distance of 0.2\,arcsec.

All these X-ray, submillimeter, and radio sources
are in the primary sample.
Except for one optically bright hard X-ray source \#62,
the others have
6.7\,$\mu$m fluxes in a range of 10--30\,$\mu$Jy.

\section{THE NATURE OF FAINT \lowercase{6.7\,$\mu$m} GALAXIES}	
\label{sect:nature}

We have identified two stars in the 6.7\,$\mu$m sample
using their image profiles in the optical
\citep{SKC+03}.
Excluding these stars from the primary sample,
we now discuss the nature of the remaining sources,
all of which are assumed to be galaxies.
We utilize the GRASIL star/dust galaxy SED model \citep{SGBD98},
which is explained in some detail in Appendix~\ref{sect:GRASIL}.

\subsection{Colors}
\label{sect:color}

\begin{figure*}
  \plotone{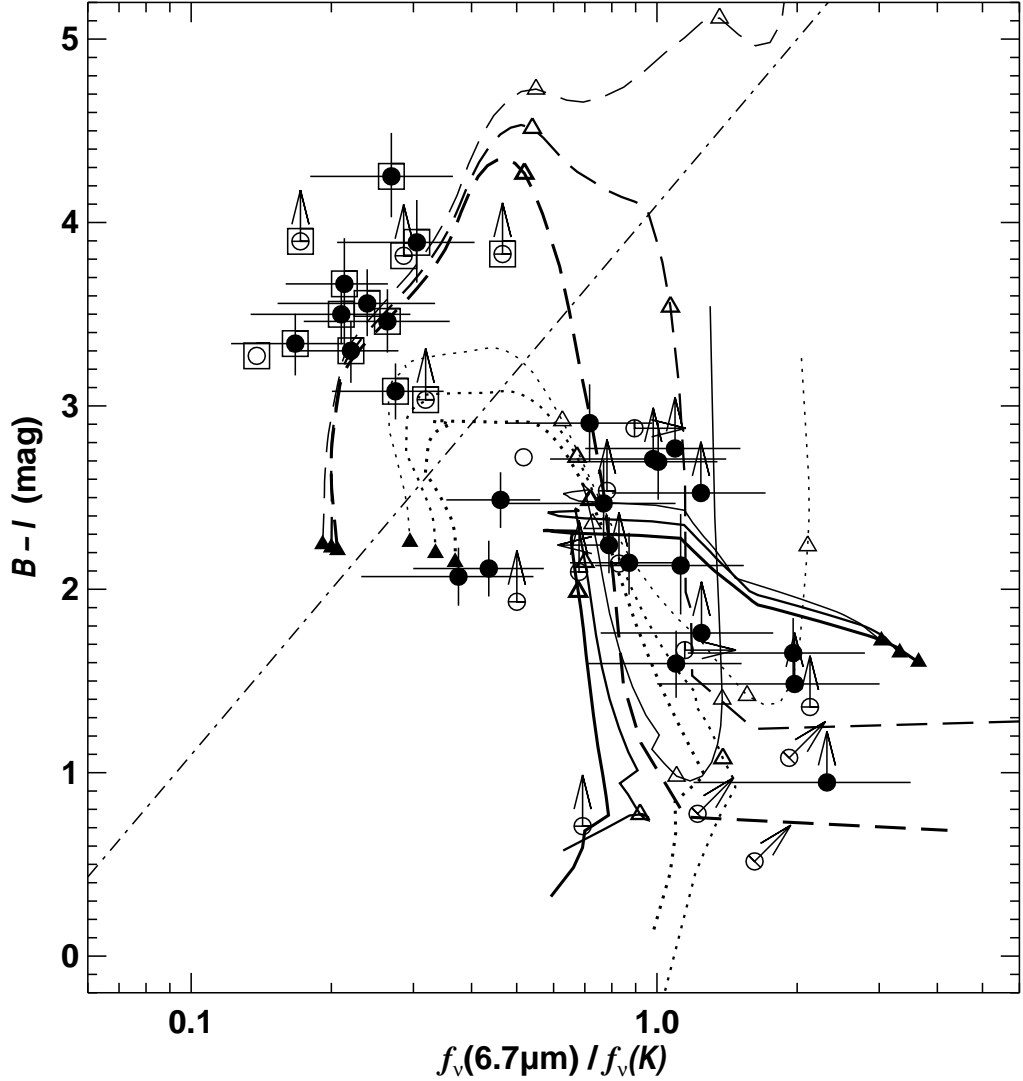}	
  \caption{
$f_{\nu}(6.7\,\mu\mathrm{m})/f_{\nu}(K)$ vs $B-I$ colors
for the identified 6.7\,$\mu$m galaxies.
The sources in the primary and supplementary samples
(Sect.~\ref{sect:samples})
are shown with filled and empty circles, respectively.
One sigma error bars are marked only for the primary sources.
For sources which have enough information,
their color limits are also indicated. 
We overlaid model predictions with the GRASIL SED library
(Appendix~\ref{sect:GRASIL}).
Galaxies with star formation histories
of the E, Sa, and Sc types (Fig.~\ref{fig:sfr}) are shown
with dashed, dotted, and solid lines, respectively.
The assumed formation redshifts are $z_\mathrm{f}=2$, 3, and 10
(thick, medium, and thin lines, respectively).
Triangles mark positions for $z=0$, 1, 2, and 3
(for $z < z_{\rm f}$; filled triangles for $z=0$).
In the upper-left region of the plot separated by the dot-dashed line,
i.e., $B-I > 3\log(f_{\nu}(6.7\,\mu\mathrm{m})/f_{\nu}(K)/2)+5$, 
we can identify a group of galaxies marked with squares (type~I).
They are consistent with the evolving E galaxies at $z=0$--1,
or the Sa galaxies at $z=0$--1 with high formation redshifts.
It should be noted
that the ages of galaxies with $z_\mathrm{f}=2$, 3, and 10
at $z=1$
are 2.5\,Gyr, 3.6\,Gyr, and 5.3\,Gyr, respectively, in our adopted cosmology.
Thus, type~I galaxies will be matured systems with old stellar populations
created by vigorous star formation episodes a long time ago
(see Fig.~\ref{fig:sfr}).
  }
\label{fig:LK_BI}
\end{figure*}

With the $K$, $I$, and $B$ band photometry,
we examined the distributions of faint 6.7\,$\mu$m galaxies
in various color-color plots.
In one plot (Fig.~\ref{fig:LK_BI}),
we identified a distinct population of faint 6.7\,$\mu$m galaxies
having red $B-I$ colors
and low ratios of $f_{\nu}(6.7\mu\mathrm{m})/f_{\nu}(K)$.
To derive ratios with 6.7\,$\mu$m fluxes,
we used zero-point fluxes of 645\,Jy, 2408\,Jy, and 3974\,Jy
for $K$, $I$, and $B$ magnitudes, respectively.
Solid circles with error bars
show sources in the primary sample.
Empty circles are for the supplementary sources.
Color limits using 3\,$\sigma$ magnitude limits
or real measurements for saturated sources
or for sources at the image boundaries
are indicated with arrows.
Some sources are not shown due to lack of information,
such as no data in the $K$ band,
or no detections in both $I$ and $B$ bands.
The concentration of galaxies at the upper-left part of the panel
is distinct from the rest of the sample by more than the relatively large errors
in $f_{\nu}(6.7\mu\mathrm{m})/f_{\nu}(K)$ ratios.
We assigned an identification flag of type~I
for galaxies in this concentration
(Table~\ref{tab:prop}).
The type~I galaxies are marked with squares
and can be separated from the rest of the galaxies
with the dot-dashed line in the plot.

On the same panel,
we overlaid some model predictions using the GRASIL SEDs
(Appendix~\ref{sect:GRASIL}).
The evolving SEDs for three $z = 0$ Hubble types
-- E, Sa, and Sc galaxies --
are shown with dashed, dotted, and solid lines, respectively.
For each Hubble type,
we assumed three formation redshifts of $z_\mathrm{f}=2$, 3, and 10
(thick, medium, and thin lines, respectively).
The locations of $z=0$ galaxies are shown with solid triangles,
while $z=1$, 2, and 3 with empty triangles (for $z < z_{\rm f}$).
We find that type~I galaxies follow
lines for the evolving E galaxies at intermediate redshifts of $z=0$--1,
regardless of their assumed formation redshifts.
The evolving Sa galaxy predictions also share the same region,
but higher formation redshifts are preferred.
In our adopted cosmology,
the ages of galaxies with $z_\mathrm{f}=2$, 3, and 10 at $z=1$
become 2.5\,Gyr, 3.6\,Gyr, and 5.3\,Gyr, respectively.
This indicates that type~I galaxies would be
old and matured systems whose stellar contents are already in place
as a consequence of vigorous star formation a long time ago
(see Fig.~\ref{fig:sfr}).

\begin{figure*}
  \plotone{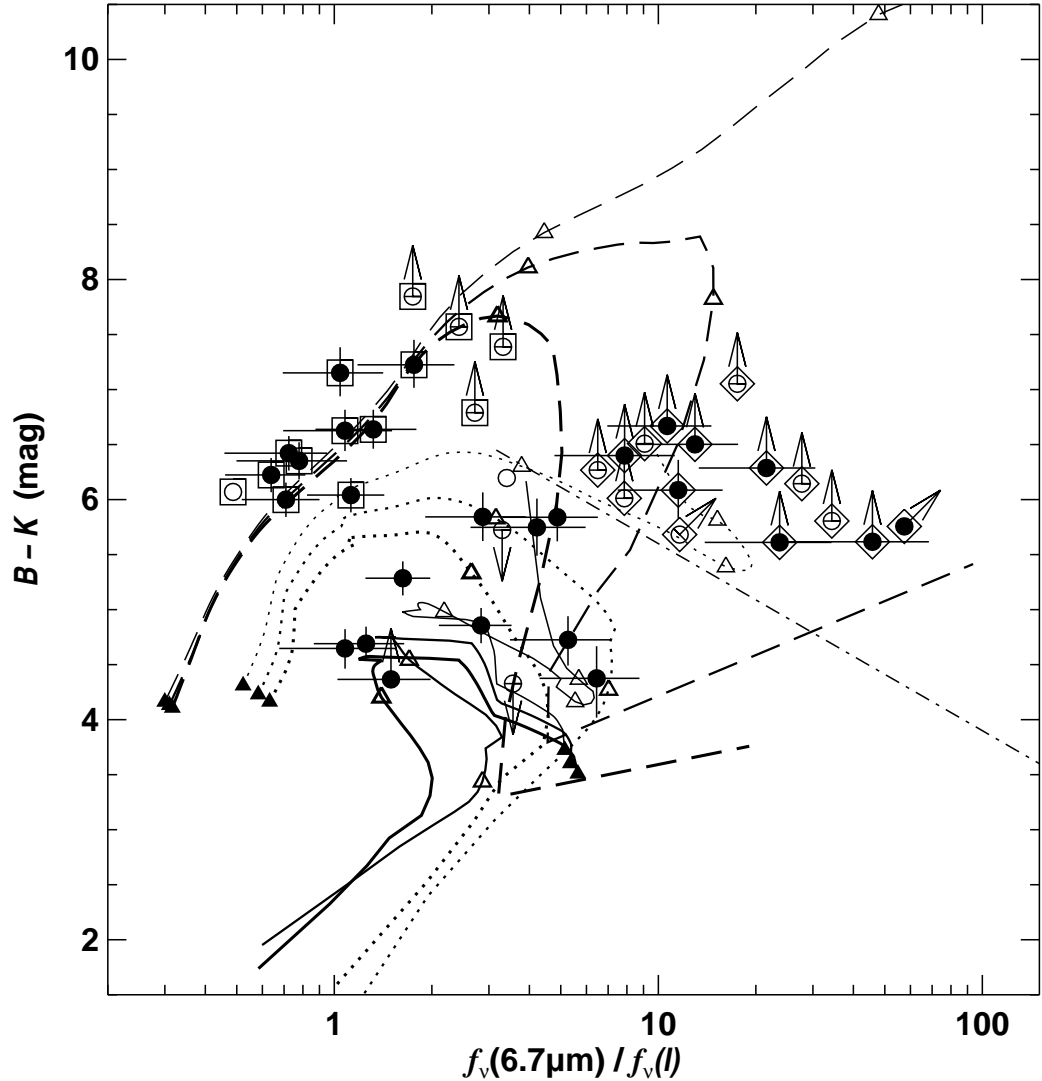}	
  \caption{
$f_{\nu}(6.7\,\mu\mathrm{m})/f_{\nu}(I)$ vs $B-K$ colors
for the identified 6.7\,$\mu$m galaxies.
Symbols are the same as in Fig.~\ref{fig:LK_BI},
although plotted sources are slightly different
due to different color sets used.
This diagram divides galaxies
in the lower-right region of Fig.~\ref{fig:LK_BI} into two types.
All the sources above the dot-dashed line,
i.e., $B-K > -1.7\log(f_{\nu}(6.7\,\mu\mathrm{m})/f_{\nu}(I))+7.3$, 
which have red colors
both in $f_{\nu}(6.7\,\mu\mathrm{m})/f_{\nu}(I)$ and $B-K$,
are marked with diamonds (type~II).
The GRASIL SED predictions show that type~II galaxies are
$z>1$ galaxies with intense star formation (evolving E or Sa galaxies).
They are consistent with being ancestors of type~I galaxies (squares).
The remaining blue galaxies are categorized as type~III.
They have the colors of the evolving Sc galaxies
or the E or Sa progenitors at $z \sim z_\mathrm{f}$,
suggesting on-going star formation.
  }
\label{fig:LI_BK}
\end{figure*}

The other galaxies in this plot follow the GRASIL predictions;
however, almost nothing can be drawn from this figure
because of the degeneracy of the model loci.
Both rapidly star-forming galaxies at high redshifts
and quiescently star-forming galaxies show
high ratios of $f_{\nu}(6.7\mu\mathrm{m})/f_{\nu}(K)$
and blue $B-I$ colors.
The color degeneracy in this part of the plot
can be disentangled to some degree by changing color combinations.
In Fig.~\ref{fig:LI_BK},
we show $f_{\nu}(6.7\mu\mathrm{m})/f_{\nu}(I)$ ratios and $B-K$ colors.
In this plot,
the evolving E or Sa galaxies can be red in both colors,
while the evolving Sc galaxies occupy the lower left part of the panel
with blue $B-K$ colors
and low ratios of $f_{\nu}(6.7\mu\mathrm{m})/f_{\nu}(I)$.
The evolving E and Sa galaxies with lower formation redshifts ($z_\mathrm{f}<3$)
can have such blue colors at their forming stage ($z \sim z_\mathrm{f}$).
Thus, these blue colors can be an indicator of on-going star formation.

The distribution of the 6.7\,$\mu$m galaxies in this plot
is rather smooth,
except for the type~I galaxies marked with squares.
However, we here introduced a separation line
(dot-dashed line)
to extract many sources not detected in the $B$ band.
These can be explained as
high redshift galaxies ($z>1$) with rapid star forming activities at the past
(evolving E or Sa galaxies).
They represent post-starburst galaxies at high redshift.
Their properties are consistent with
those of ancestors of type~I galaxies.
We here categorize them as type~II and mark them with diamonds.
The remaining galaxies are assigned as type~III.
They are on-going star formers at $z=0$--2,
a combination of mildly star forming galaxies
and vigorously star forming galaxies with low formation redshifts.

The division of the sources into types I, II, and III
is made in color-color plots
with photometry in all four bands -- 6.7\,$\mu$m, $K$, $I$ and $B$.
We find that this division can also be applied to the color-color plot of
$f_{\nu}(6.7\mu\mathrm{m})/f_{\nu}(I)$
vs $f_{\nu}(6.7\mu\mathrm{m})/f_{\nu}(B)$.
In this plot,
we can assign type identifiers to all the 6.7\,$\mu$m sources,
with the help of the distribution of the sources
whose types have been defined in the previous two plots.
Even unidentified sources
could have a nominal type of II
by their red colors
due to their flux limits at the $I$ and $B$ bands.
For the primary sample,
the number ratio for the 31 galaxies becomes
type $\mathrm{I} : \mathrm{II} : \mathrm{III} = 11 : 10 : 10$
including the unidentified source \#27 with type~II
(Table~\ref{tab:prop}).
This almost even distribution skews toward
the dominance of type~II in the supplementary sample.
For the 32 sources in this sample,
the ratio becomes
type $\mathrm{I} : \mathrm{II} : \mathrm{III} = 5 : 23 : 4$.
The dominance of type~II in this fainter sample is still valid
even if the seven unidentified sources assigned type~II are excluded.

\subsection{Redshifts}

In order to obtain absolute quantities such as stellar masses
for the 6.7\,$\mu$m galaxies, we need their redshifts.
For the 55 identified 6.7\,$\mu$m galaxies,
we do have 21 spectroscopic redshifts $z_\mathrm{sp}$,
of which 18 are for the primary sample
\citep{SCHG94,CSHC96,BCMR01}.
For others,
we derived photometric redshifts $z_\mathrm{ph}$
in the following way.

\subsubsection{Photometric redshifts}
\label{sect:zph}

Photometric redshifts were estimated
by minimizing $\chi^2$ values
between photometric measurements and model estimates.
This is basically the same method adopted in the the $hyperz$ code
by \citet{BMP00}.
We used 6.7\,$\mu$m, $K$, $I$, and $B$ band fluxes or their limits
listed in Table~\ref{tab:prop}.
For sources in Table~\ref{tab:xsr}, we tried to utilize
their flux values or their limits
at the X-ray, submillimeter, and radio wavelengths as well.
Model estimates were obtained
using two sets of the star/dust SEDs in the GRASIL library;
the evolving SEDs and the local ones
(Appendix~\ref{sect:GRASIL}).
We also added two SEDs expanded to the X-ray wavelengths;
a NGC\,6240 SED compiled by \citet{H00}
and a GRASIL Arp220 SED
complemented with X-ray and radio observations
\citep{I99,CY00}.

The $\chi^2$ minimization was executed SED by SED in our code.
The SEDs used have a long wavelength baseline;
the GRASIL local SEDs from UV to submillimeter,
the GRASIL evolving SEDs from UV to radio,
and the two additional SEDs from X-ray to radio.
They are likely to have significance variations at different wavelengths;
however, we neglected such errors in the SEDs.
We thus considered noise only in the absolute photometric measurements.
For each SED, we calculated $\chi^2$ values
at 100 redshifts from $z=0.1$ to 10.
The interval of the redshift grid was constant in $\log (1+z)$.
Once the redshift giving the minimum $\chi^2$ value was derived,
we repeated the calculation around this redshift
with a ten times finer redshift resolution.
After completing the $\chi^2$ minimization for all the SEDs,
we determined a photometric redshift giving the minimum $\chi^2$ value
among the SEDs.
We also found that
in most cases,
the redshift giving the minimum $\chi^2$ value
for each SED is consistent with the photometric redshift
determined by the global minimum in $\chi^2$.
It should be noted that
the final photometric redshifts were all determined
with the UV-to-submillimeter SEDs
because the SEDs expanded to the radio and/or X-ray
resulted in high $\chi^2$ values.
For the comparison of $\chi^2$ values
among SEDs with different wavelength spans,
we took proper account of both the
values of $\chi^2$
and the number of the detected data points used to calculate $\chi^2$.
In these calculations,
we did not require that the age of the SED template
be less than the age of the universe at a given redshift.
Previous works on photometric redshifts
have given good results with local templates at high redshifts
\citep[see e.g.,][]{H98}.
We therefore tried to fit local SEDs even at $z=10$.
Old evolving SEDs were also treated as local SEDs.
This assures a possible range of SED variations.
Actually, we found that the results of this approach
admitting the age-inconsistency in the fits
gave the best results
in the comparison with spectroscopic redshifts explained below.

\begin{figure*}
  \plotone{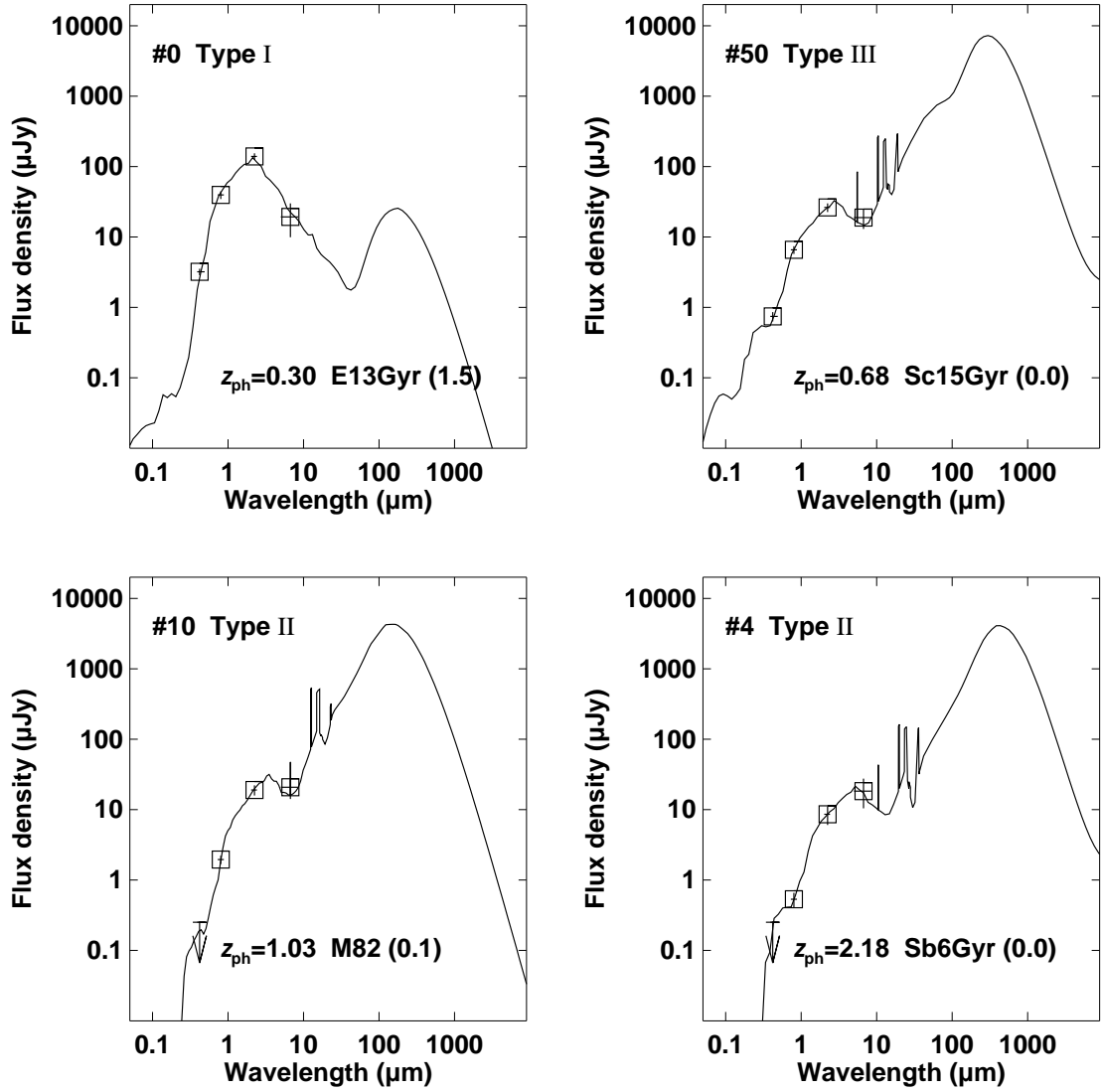}	
  \caption{
Some examples of the SED fits used to derive photometric redshifts.
Detections are shown with squares,
in which flux errors and band widths are indicated
with vertical and horizontal bars.
Arrows are $3\,\sigma$ upper limits.
Source names and their photometric redshifts $z_{\rm ph}$ are given
with their color types, GRASIL SED names,
and $\chi^2$ values (in parentheses).
  }
\label{fig:zfit}
\end{figure*}

We show some fitting results at different redshifts
in Fig.~\ref{fig:zfit}.
Photometric redshifts and their 90\,\% confidence limits
\citep{A76}
are listed in Table~\ref{tab:prop}.
The names of the best-fit SEDs, their $\chi^2$ values,
and some ancillary information of the fits are also shown.
It should be noted that many other SEDs gave quite similar redshift estimates
with somewhat larger $\chi^2$ values.
For source \#38, which has the largest $\chi^2$ value in the sample,
12 SEDs gave $z=0.74$--0.98 with $\chi^2=9.9$--15.
On the other hand,
many sources have $\chi^2=0.0$, though they are truncated values.
This indicates overestimates of our photometric errors.
This is likely in our case
due to quadratic summation
of error components,
each of which was difficult to determine independently.
Our data points are very sparse in wavelength.
In a sense, we might have used too many sets of SEDs for such data sets.
It can be said that we
have allowed the SEDs to shift freely in redshift space
to take account of the symmetric distribution of photometric errors.
We have succeeded in eliminating outlier fits;
however, some fits could have
fallen at rather artificial $\chi^2$ minima.
Thus, the confidential intervals are more reliable
than photometric redshift values themselves.
It should be noted that
the photometric redshift determination technique
basically utilizes the characteristic spectral features common to all SEDs.
There is some information in the fit; however,
the very particular SED names
identified as giving the best photometric redshifts
do not exclude other SED types.

\begin{figure*}
  \plotone{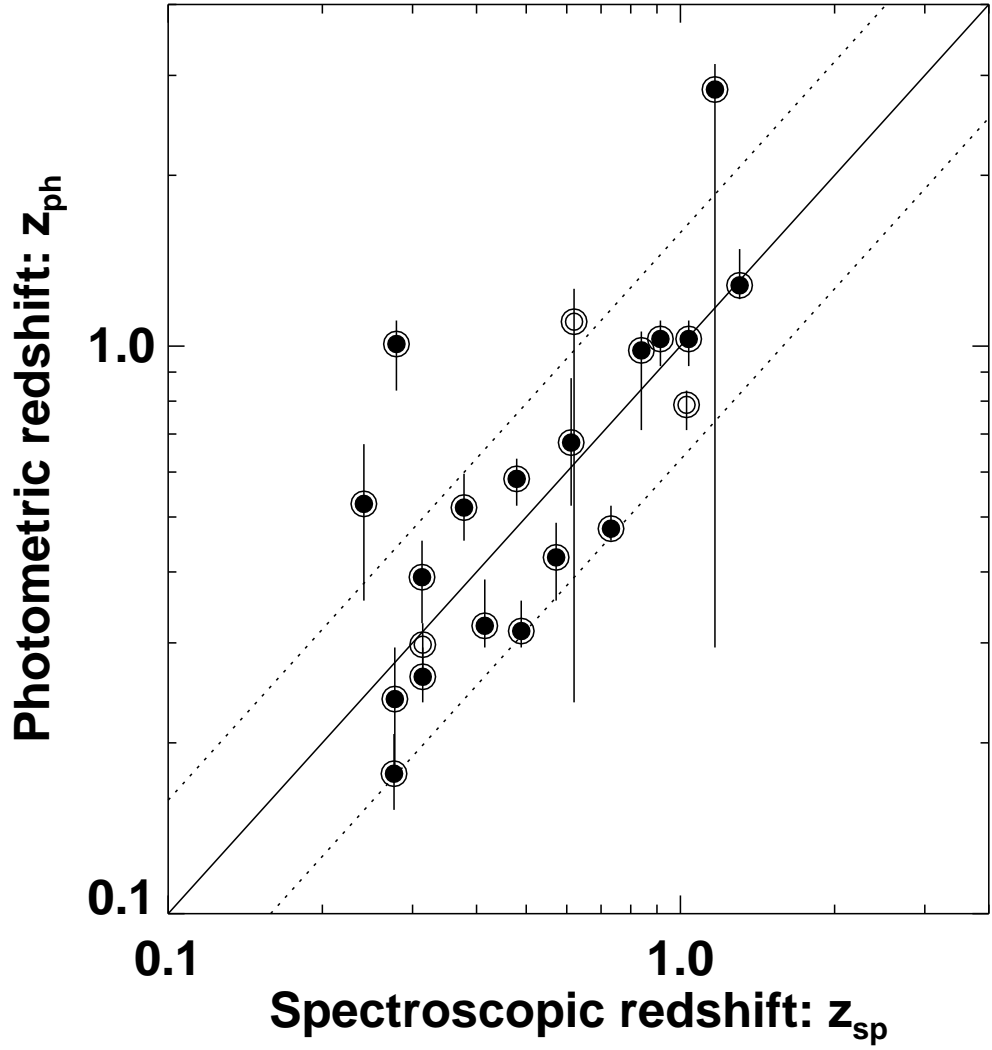}	
  \caption{
Comparison of photometric redshift estimates to
spectroscopic redshift measurements.
The photometric redshifts were derived as in Sect.~\ref{sect:zph}
and their nominal 90\,\% confidence limits are shown with vertical bars.
Almost all the estimates are within one sigma ranges (0.2\,dex; dotted lines)
from the unity relation (solid line).
The primary and supplementary sample galaxies
(Sect.~\ref{sect:samples})
are shown with filled and empty symbols, respectively.
  }
\label{fig:zsp_zph}
\end{figure*}

The reliability of these redshift estimates can be addressed
by computing photometric redshifts
for sources with spectroscopic redshifts.
The results are shown in Fig.~\ref{fig:zsp_zph}.
Of 21 sources with spectroscopic redshifts,
10 photometric redshifts were derived with the age-inconsistent
evolving SEDs,
6 with local templates, and 5 with age-consistent SEDs.
This result assures the usefulness of the admittance of age-inconsistency
in the fits.
Actually, most of the photometric estimates follow the identity relation
with their spectroscopic measurements.
The nominal dispersion was 0.2\,dex in $\log z$
(dotted lines).
Only four sources, \#62, \#15, \#29, and \#16
with spectroscopic redshifts of $z \sim 0.2$, 0.3, 0.6, and 1.2, respectively,
gave photometric redshifts larger by more than one sigma.
However, their large 90\,\% confidence intervals
indicate that they are essentially problematic data sets.
For example,
source \#62 lacks any flux constraint in the $K$ band.

The four photometric bands, $B$, $I$, $K$, and 6.7\,$\mu$m,
are widely and evenly distributed in wavelength
(Fig.~\ref{fig:zfit}).
The determination of photometric redshifts with these data sets
appears to depend on a characteristic spectral peak around 1.6\,$\mu$m
corresponding to the H$^-$ opacity minimum for the stellar continuum emission.
The use of this 1.6\,$\mu$m bump for photometric redshifts
with photometric data sets without wavelength gaps
is described in \citet{S02}.
In wider wavelengths,
the spectral slope is generally monotonic below 1.6\,$\mu$m,
while there could be UIB emission longward of the 1.6\,$\mu$m peak
until 6.7\,$\mu$m
(i.e. 3.3\,$\mu$m and 6.2\,$\mu$m).
There is a possibility that
some UIB emission was erroneously treated as a shorter wavelength feature.
Misidentification of 6.2\,$\mu$m emission with a 3.3\,$\mu$m or 1.6\,$\mu$m
feature gives a wrong redshift of $z \sim 1$ or 3
and that of 3.3\,$\mu$m emission with a 1.6\,$\mu$m feature
gives $z \sim 1$.
Unfortunately,
the 6.7\,$\mu$m photometric band is very broad (5--8.5\,$\mu$m)
and the gaps between the four photometric bands are somewhat large.
Some sources are even lacking some of the four flux values.
This all could give a reason for the four high redshift outliers
in Fig.~\ref{fig:zsp_zph}.
The effects of $z_{\rm ph} > z_{\rm sp}$ estimates
will be addressed in the following sections.

\subsubsection{Fluxes and redshifts}
\label{sect:z_flux}

With redshifts determined spectroscopically or photometrically,
6.7\,$\mu$m fluxes for all 55 identified galaxies
are plotted as a function of redshift in Fig.~\ref{fig:z_flux}.
Flux errors are indicated with vertical bars,
while horizontal bars are 90\,\% confidence limits for photometric redshifts.
Sources with spectroscopic redshifts are marked with double circles.
Galaxies in the primary sample are highlighted with solid symbols.

For the primary sample,
fainter galaxies tend to have higher redshifts.
This usual trend seems to weaken below a flux level of 30\,$\mu$Jy, where
high redshift galaxies start to appear regardless of their flux values.
This first finding is a direct result of the substantial depth of our imaging.
Some photometric redshifts could be erroneously high
(Sect.~\ref{sect:zph}),
but the high redshift tail
can be recognized even within the spectroscopic primary sample alone.
There are some $z_\mathrm{sp}>1$ galaxies with fluxes of $\sim 20\,\mu$Jy.
Most of the supplementary sample have $z_\mathrm{ph}>1$.
Because they are generally fainter than 30\,$\mu$Jy,
this behavior itself strengthen the high redshift tail
seen in the primary sample.
Two bright supplementary sources at $z \sim 2$ (\#12 and \#34)
are detected in relatively high noise regions and lack $K$ data.
Thus, their significance is taken to be low.

In this plot, model predictions are overlaid using the GRASIL library
(Appendix~\ref{sect:GRASIL}).
Dashed, dotted, and solid lines are evolving E, Sa, and Sc galaxies
and thick, medium, and thin lines are
cases with formation redshifts of $z_\mathrm{f}=2$, 3, and 10.
The distribution of the 6.7\,$\mu$m galaxies
is almost bracketed by these model predictions.
At low redshifts ($z \sim 0.3$),
most of the 6.7\,$\mu$m galaxies are consistent with
the evolving Sa and Sc galaxy models.
The steep flux-redshift slopes for the evolving Sc galaxies
are due to the effects of strong dust emission in the Sc galaxies
to the 6.7\,$\mu$m observing band.
At high redshifts ($z \sim 1$),
the distributions of both primary and supplementary galaxies
are centered at model predictions for the evolving Sa galaxies.
The flat flux-redshift distribution of the 6.7\,$\mu$m galaxies
at high redshifts can be reproduced well
by a factor of a few less luminous version of the evolving E galaxy models.
Although the GRASIL models cannot be arbitrarily rescaled,
especially at the UV and far-infrared wavelengths
where effects of dust are significant
(Appendix~\ref{sect:GRASIL}),
these model loci at $z>1$ are primarily
due to stellar emission in the rest-frame near-infrared,
which accepts rescaling.

We mark the locations of the X-ray, submillimeter, and radio sources
with letters, 'X', 'S', and 'R', respectively.
Except for one optically bright hard X-ray source (\#62),
all are within the high redshift tail.

\subsection{Comparison with a $K<20$ Sample}

Some of the characteristics of the faint 6.7\,$\mu$m galaxies
can be assessed in a comparison with a $K$ band magnitude limited sample.
We here utilize the $K<20$ spectroscopic sample in the SSA13 field
presented in \citet{CSHC96}.
The field coverage of this $K$ sample is a few times larger than
that of the ISOCAM map.

\begin{figure*}
  \plotone{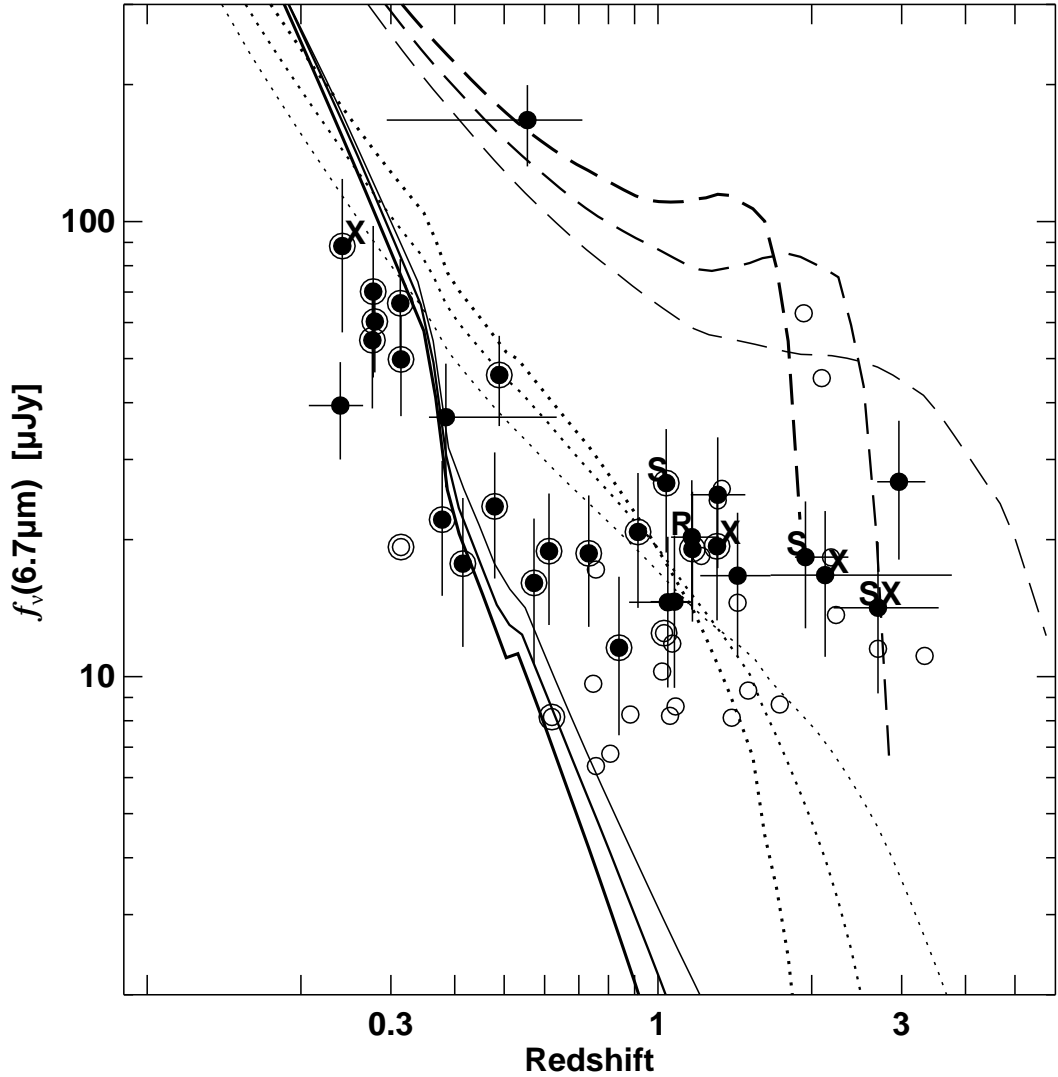}	
  \caption{
6.7\,$\mu$m fluxes as a function of redshift.
Sources with spectroscopic redshifts are indicated with double circles.
For photometric redshifts,
their 90\,\% confidence limits are indicated with horizontal bars.
Vertical bars are one sigma flux errors.
Galaxies in the primary sample are shown with filled circles,
while the supplementary sources are marked with empty circles
with no error bars.
Letters, 'X', 'S', and 'R' indicate
the X-ray, submillimeter, and radio counterparts
in Table~\ref{tab:xsr}, respectively.
Dashed, dotted, and solid lines
are model predictions for the GRASIL evolving E, Sa, and Sc galaxies, respectively
(Appendix~\ref{sect:GRASIL}).
For each SED,
formation redshifts are assumed to be $z_\mathrm{f}=2$, 3, and 10
(thick, medium, and thin lines, respectively).
  }
\label{fig:z_flux}
\end{figure*}

\begin{figure*}
  \plotone{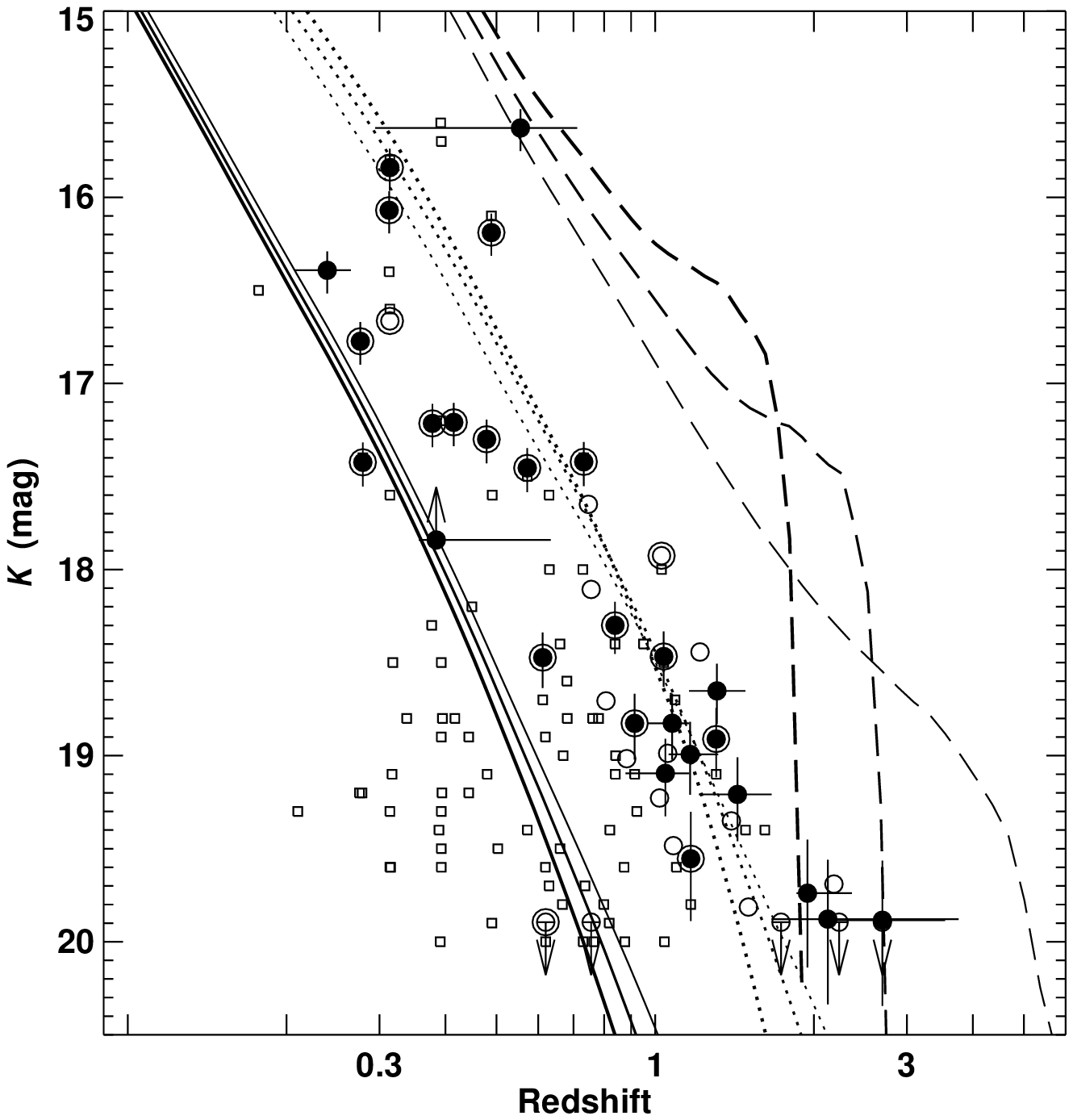}	
  \caption{
$K$ magnitudes of the 6.7\,$\mu$m galaxies.
Symbols are the same as in Fig.~\ref{fig:z_flux}.
Some of the 6.7\,$\mu$m sources
are shown with their upper or lower magnitude limits,
or not shown due to the lack of their $K$ photometry
(Table~\ref{tab:prop}).
For reference,
a $K>20$ spectroscopic sample \citep{CSHC96} is overlaid with small squares.
The 6.7\,$\mu$m-selected sample does not include
faint $K$ sources at low redshifts,
some of which are even fainter than the GRASIL evolving Sc galaxy predictions.
  }
\label{fig:z_K}
\end{figure*}

First, we show a $K-z$ diagram
both for the 6.7\,$\mu$m- and $K$-selected samples
in Fig.~\ref{fig:z_K}.
Symbols are the same as in Fig.~\ref{fig:z_flux},
except $K<20$ galaxies are overlaid with small squares.
Note that some 6.7\,$\mu$m galaxies plotted in Fig.~\ref{fig:z_flux}
are not shown here because they lack $K$ photometry.
6.7\,$\mu$m galaxies with fainter $K$ magnitudes
tend to have larger redshifts.
This feature is well described with the GRASIL evolving Sa galaxy model,
and the distribution of the 6.7\,$\mu$m galaxies in the plot
lies between the evolving E and Sc galaxy predictions.
No significant difference is seen between
the primary and supplementary samples.

In contrast,
the distribution of the $K$ band selected galaxies
extends to a region fainter than the evolving Sc galaxy model predictions
and contrasts with the 6.7\,$\mu$m sample.
The 6.7\,$\mu$m sample
preferentially selects
a high redshift population at each $K$ magnitude.
At intermediate redshifts ($z=0$--1),
$K$ band emission is a good indicator of stellar masses
(Sect.~\ref{sect:SMLR}).
The evolving Sc galaxy models at $z \sim 0.4$
have stellar masses of 0.1--0.2\,$M_\mathrm{star}^*$,
where we adopt $M_\mathrm{star}^*=1.7 \times 10^{11}$\,M$_\odot$
\citep{CNB+01}.
Thus, the $z<1$ faint $K$ galaxies
should have very small stellar masses.

\begin{figure*}
  \plotone{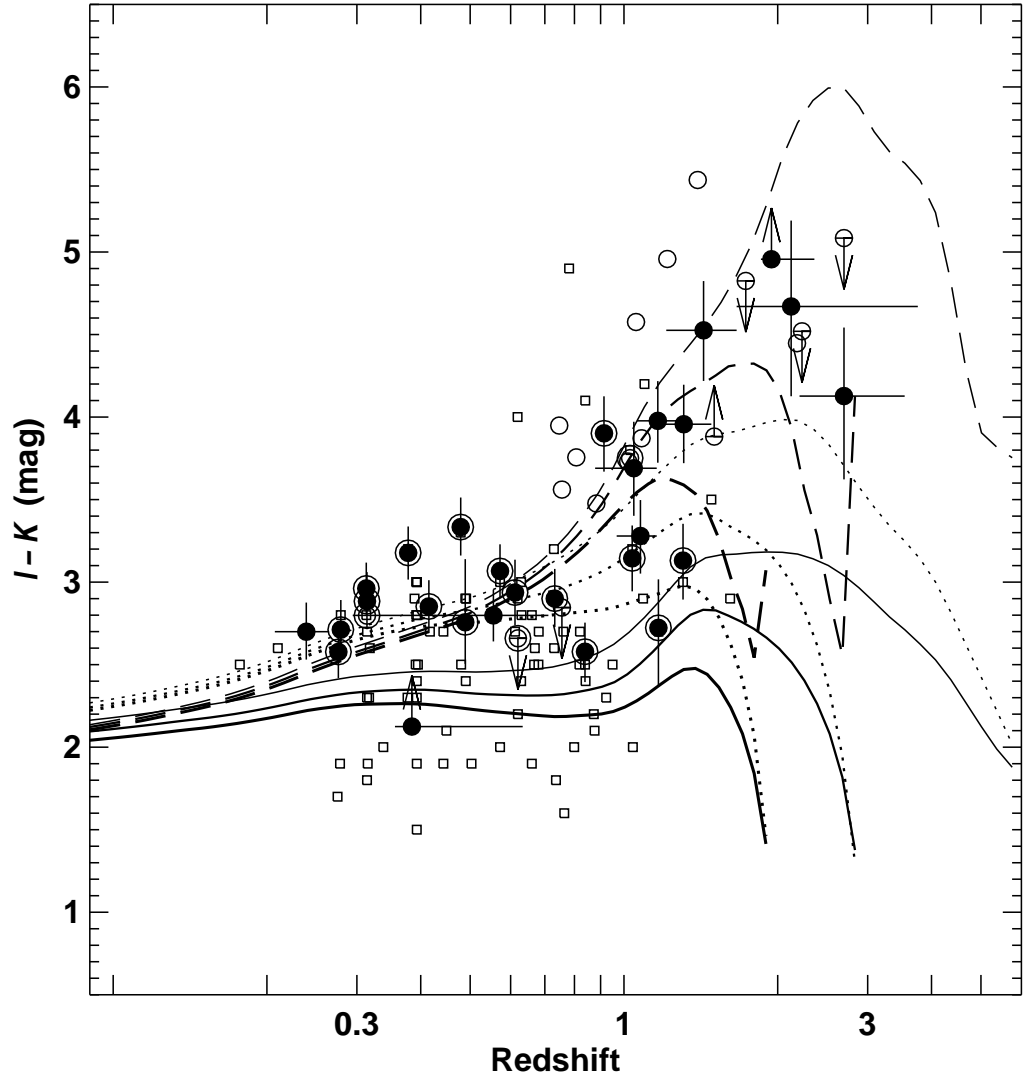}	
  \caption{
$I-K$ colors of the 6.7\,$\mu$m galaxies.
Symbols are the same as in Fig.~\ref{fig:z_K}.
The 6.7\,$\mu$m-selected sample does not include
blue $I-K$ sources at low redshifts,
some of which are even bluer than the GRASIL evolving Sc galaxy predictions.
  }
\label{fig:z_IK}
\end{figure*}

Next, we show $I-K$ colors
as a function of redshift in Fig.~\ref{fig:z_IK}.
The 6.7\,$\mu$m galaxies have red $I-K$ colors
especially at $z>1$.
This reddening trend is expected with the GRASIL models.
However, some 6.7\,$\mu$m galaxies exceed the red envelope
of the GRASIL predictions.
These excess red $I-K$ colors are seen at intermediate redshifts
for sources in the primary sample, and seem to be consistent
with the presence of some high redshift ($z>1$) galaxies
in the supplementary sample.
The very red $I-K$ colors at $z>1$
indicate rapid star formation at very early epochs
in the context of the GRASIL model.
The reddest $I-K$ colors in the 6.7\,$\mu$m sample
are comparable to those of an evolving E galaxy with a formation redshift
of $z_\mathrm{f}=10$.
For a brighter 6.7\,$\mu$m sample,
the existence of red $I-K$ sources is also reported
in \citet{FHD+99}.

Most of the 6.7\,$\mu$m galaxies have $I-K$ colors
redder than the evolving Sc galaxy predictions.
However, some $K$ band selected galaxies have blue colors such as $I-K<2$.
Their blue colors imply that young stellar populations are dominant.
As indicated with the GRASIL predictions at $z \sim z_\mathrm{f}$,
very blue $I-K$ colors can only be seen in the forming stages of galaxies.
Thus,
the blue $K$-selected galaxies should experience
star forming activities less burst-like than the evolving  Sc galaxies,
or they should start to form stars in a wide range of redshift at $z=0.3$--1.

We find that the $K$ band galaxies bluer than the evolving Sc galaxy predictions
have $K$ magnitudes fainter than the Sc galaxy models
(Fig.~\ref{fig:z_K}).
These $K$ selected galaxies having blue $I-K$ colors and faint $K$ magnitudes
at intermediate redshifts ($z<1$)
should have small stellar masses at their forming stages.
Faint $K$ band selected samples
will be contaminated with this population of young dwarf galaxies
at low redshifts.
However, faint 6.7\,$\mu$m selected samples will be free from it.
It means that an efficient search for high redshift galaxies
will be achieved with deep surveys at the mid-infrared, e.g., 6.7\,$\mu$m.
  
\section{STELLAR MASS}
\label{sect:SM}

\subsection{Stellar Mass-to-Light Ratios}
\label{sect:SMLR}

For the conversion of observed luminosities
to their stellar masses,
we calculated stellar mass-to-light ratios
at the observed bandpasses.
These ratios are to be applied to the observed values directly.
This approach is different from others in
literature (e.g., taking the observed $K$ band
flux for a $z=1$ galaxy,
converting it to the rest-frame $K$ band with an assumed
SED, and applying a stellar mass-to-light ratio at the rest-frame $K$ band).
We eliminated such conversion to a fixed bandpass.

We calculated both stellar mass $M_\mathrm{star}$
and an in-band luminosity $L$
that should be observed directly
as a function of redshift.
The in-band luminosities are presented
in units of the bolometric luminosity of the Sun
(1\,L$_\odot^\mathrm{bol}=3.85 \times 10^{26}$\,[W]).
We adopted the evolving SEDs in the GRASIL library
(Appendix~\ref{sect:GRASIL})
and the resulting stellar mass-to-light ratios $M_\mathrm{star}/L$
are shown for the $B$, $I$, $K$, and ISOCAM LW2 bands
(Fig.~\ref{fig:z_ml4}).
Dashed, dotted, dot-dashed, and solid lines represent
the evolving E, Sa, Sb, and Sc galaxies.
Formation redshifts were assumed to be $z_\mathrm{f}=5$ and 10
(thick and thin lines, respectively).

It is well known
that the rest-frame $K$ band light is a good indicator of stellar mass.
This fact is reflected in the $K$ band panel,
showing very small dispersions among the cases for different galaxies
or different formation redshifts,
especially at low redshifts.
At low redshifts,
the dispersions among different models
become larger at shorter wavelengths, here the $I$ and $B$ bands.
The large dispersions at low redshifts in the 6.7\,$\mu$m panel
are due to the contamination from dust emission.

Moving to larger redshifts,
the situation regarding the dispersion among star-forming histories
will change among the different observing bands.
Very roughly,
the effective wavelengths of the $B$, $I$, $K$, and LW2 bands
are two time larger than those of the neighboring shorter bands.
The dispersions for the $I$, $K$, and LW2 observing bands at $z \sim 1$
are very similar to those for the $B$, $I$, and $K$ bands at $z \sim 0$.
This means that
6.7\,$\mu$m light will become a better stellar mass indicator
at high redshifts.
On the contrary, $B$ band fluxes cannot be used to estimate stellar masses
at high redshifts.
They correspond to rest-frame UV fluxes,
which are good indicators of star formation rates.
Emission at short wavelengths such as UV emission
is sensitive to ages of galaxies, or formation redshifts.
Such effects are actually seen in the $B$ band panel.
They can be seen in the $K$ band panel at high redshifts.

\subsection{Stellar Masses of Faint 6.7\,$\mu$m Galaxies}
\label{sect:SM6.7}

Stellar masses of the faint 6.7\,$\mu$m galaxies were derived
from rest-frame near-infrared light
taking into account the behavior of the stellar mass-to-light ratios
in Fig.~\ref{fig:z_ml4}.
Namely, the conversion of luminosities to stellar masses
was performed using 6.7\,$\mu$m luminosities for sources at $z>1$
and $K$ band luminosities for sources at $z<1$.
For low redshift galaxies with no $K$ photometry,
we used $I$ band luminosities instead.
This approach minimized the uncertainties in stellar mass-to-light ratios.
In addition,
we determined the SED to derive stellar mass-to-light ratios
by using all the photometry data from UV to submillimeter
via $\chi^2$ minimization to the evolving SEDs in the GRASIL library
(Appendix~\ref{sect:GRASIL}).
Because older SEDs usually have higher stellar mass-to-light ratios,
we set an age constraint for the fitting template.
No SEDs older than the age of the universe at the source's redshift
were used for the fitting.
This will avoid artificially high estimates of stellar mass-to-light ratios
at the expense of increasing the nominal $\chi^2$ of the best-fit.
Note that $\chi^2$ values for sources with spectroscopic redshifts
can be larger than those for sources with photometric redshifts.
This is because some fraction of inappropriateness of the model SEDs
is absorbed in the fits in determining photometric redshifts.
However, such errors are
reflected in the uncertainties of the photometric redshifts.

\begin{figure*}
  \plotone{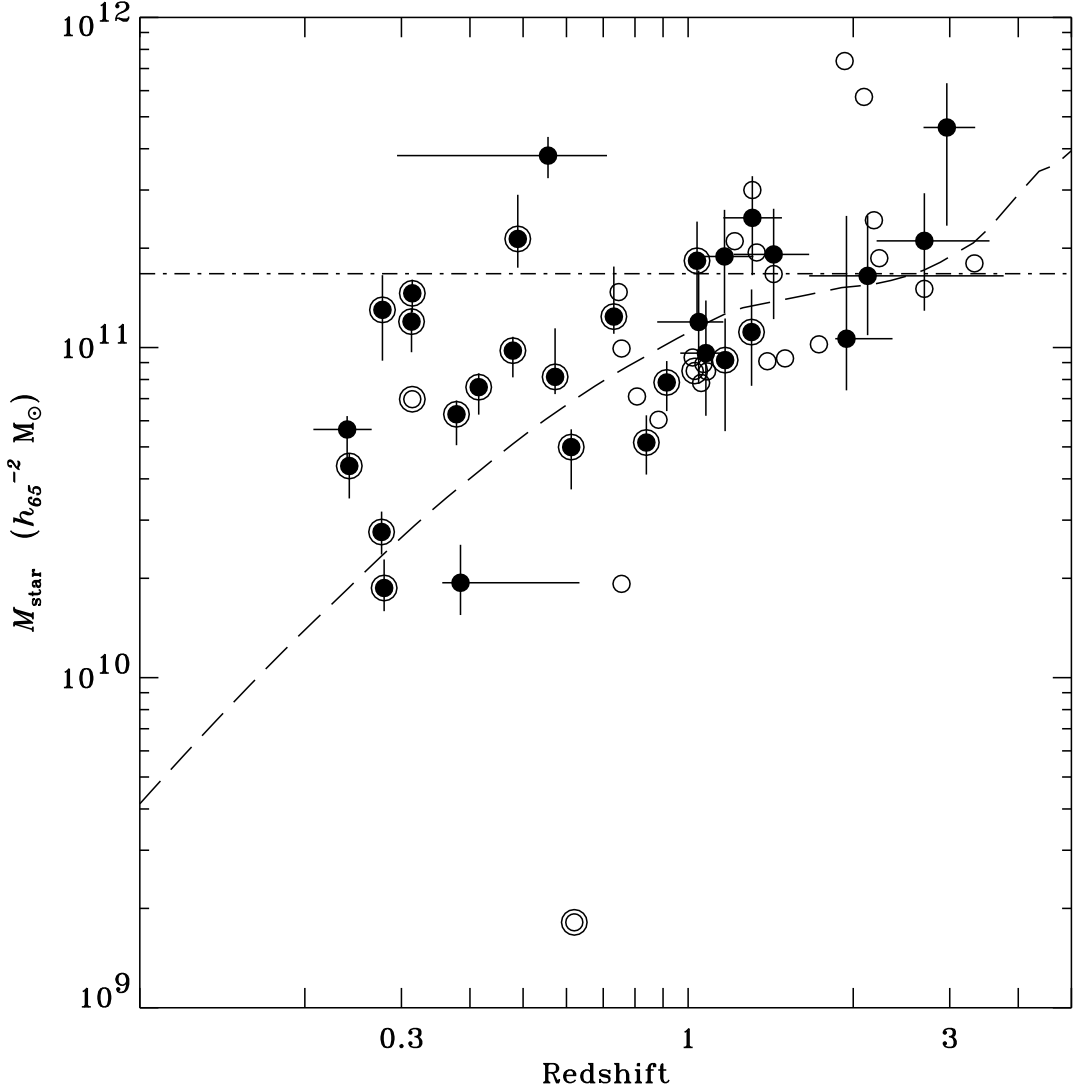}	
  \caption{
Stellar masses of the 6.7\,$\mu$m galaxies.
Symbols are the same as in Fig.~\ref{fig:z_flux}.
The stellar masses are
derived from 6.7\,$\mu$m luminosities at $z>1$
or $K$ band luminosities otherwise.
If no $K$ band luminosities are available,
$I$ band luminosities are used instead.
Note that errors in stellar mass
come mainly from errors in photometry,
not from errors in stellar mass-to-light ratios
derived from the SED fit.
A horizontal dot-dashed line indicates
a characteristic stellar mass of local galaxies
\citep{CNB+01}.
A dashed line shows stellar mass predictions
assuming the GRASIL evolving E galaxy
with a formation redshift of $z_\mathrm{f}=10$.
They are scaled to
the minimum 6.7\,$\mu$m flux of the primary sample (12\,$\mu$Jy).
This curve itself is useful
to indicate the redshift dependence of the detection limit
especially at the high redshift end
where the scaling from the GRASIL model is almost unity.
However, the scaling factor becomes very small,
$\sim 0.03$ (0.1\,$M_\mathrm{star}^*$) at $z=0.2$
(Appendix~\ref{sect:GRASIL}).
So, the uncertainties of the dashed line at the low redshift will be large.
Moreover, 6.7\,$\mu$m fluxes start to be affected by dust emission
at such low redshifts.
Note that stellar masses were derived from $K$ fluxes at $z<1$.
  }
\label{fig:z_m}
\end{figure*}

The resulting stellar masses
using the best-fit stellar mass-to-light ratios
are listed in Table~\ref{tab:prop}.
The names of the SEDs used, their $\chi^2$ values,
and some ancillary information of the fitting are also shown.
In order to derive errors in stellar mass-to-light ratios,
we searched SED fits with $\Delta \chi^2=1$
from the best-fit SED
\citep{A76}.
For some sources,
no such SED fits were found to determine confidence limits
because of the sparse grid of SEDs.
In such cases, we adopted stellar mass-to-light ratios
of the SED fits with the minimum $\Delta \chi^2$ values.
Even with such overestimated noise,
we found that the photometric errors
dominate the total uncertainties in the stellar masses.
This is because
any effects on the stellar mass-to-light ratio values
were minimized as a result of our hybrid conversions
using 6.7\,$\mu$m and $K$ band luminosities.
The actual trend can be seen in Fig.~\ref{fig:z_m}.
Larger errors in stellar masses for sources at $z>1$
originate in larger photometric errors in their 6.7\,$\mu$m fluxes.
It is also noted that
our stellar mass estimates are not likely to be affected by
the limitation of rescaling in the GRASIL SED model
because the rest-frame near-infrared light
is negligibly affected by dust
(Appendix~\ref{sect:GRASIL}).

Many 6.7\,$\mu$m galaxies have stellar masses
as massive as $10^{11}$\,M$_\odot$.
The characteristic stellar mass of local galaxies
$M_\mathrm{star}^*=1.7 \times 10^{11}$\,M$_\odot$
\citep{CNB+01}
is indicated with a horizontal dot-dashed line.
Also shown (dashed line) are stellar masses of the evolving E galaxy
with a formation redshift of $z_\mathrm{f}=10$
scaled to the minimum 6.7\,$\mu$m flux of the primary sample (12\,$\mu$Jy).
This shows that $M_\mathrm{star}^*$ galaxies could be detected out to $z \sim 3$,
if such galaxies were present.
Some 6.7\,$\mu$m sources have stellar masses
smaller than this GRASIL prediction.
They are likely to have star formation histories
less burst-like than the evolving E galaxy
or lower formation redshifts of $z_\mathrm{f}<10$.
This curve itself is useful
to indicate the redshift dependence of the detection limit
especially at the high redshift end
where the scaling from the GRASIL model is almost unity.
However, the scaling factor becomes very small,
$\sim 0.03$ (0.1\,$M_\mathrm{star}^*$) at $z=0.2$
(Appendix~\ref{sect:GRASIL}).
So, the uncertainties of the dashed line at low redshift will be large.
Moreover, 6.7\,$\mu$m fluxes start to be affected by dust emission
at such low redshifts.
Note that stellar masses were derived from $K$ fluxes at $z<1$.

The distribution of the supplementary sample
is similar to that of the primary sample.
Two of the highest stellar mass sources in the supplementary sample
at $z \sim 2$
(sources \#12 and \#34)
are the dubious sources mentioned in Sect.~\ref{sect:z_flux}.
The lowest stellar mass for the supplementary sample
at $z \sim 0.6$ (source \#29)
falls somewhat far from the distribution of other sources.
This is derived from its $I$ band luminosity
and could be somewhat less secure.

\subsection{Epoch Dependent Stellar Mass Functions}
\label{sect:SMF}

Here we examine the redshift dependence of the stellar mass function.
The epoch-dependent stellar mass functions $\Phi(M_\mathrm{star},z)$
were estimated
with the 1/$V_\mathrm{max}$ method
\citep[e.g.,][]{TYI00}
as
\begin{equation}
\Phi(M_\mathrm{star},z)=\sum \frac{1}{V_\mathrm{max}}.
\end{equation}
No sources in the supplementary sample
were used in the summation
in order to avoid the unwanted effects of spurious detections
(Sect.~\ref{sect:samples}).
Taking account of errors in photometric redshifts,
we adopted three broad redshift bins;
$z=0.2$--0.5, 0.5--1.2, and 1.2--3.0.
The widths of these bins are almost 0.4\,dex in $\log z$.
The maximum volume $V_\mathrm{max}$ for each source was derived according to
\begin{equation}
V_\mathrm{max}=\int_{z_\mathrm{min}}^{z_\mathrm{max}}
\Omega(<S(z)) \frac{d^2V}{d\Omega dz} dz,
\end{equation}
where $\Omega(<S)$ is the effective solid angle in this survey
for a source with a flux of $S$.
This function is presented in \citet{SKC+03}.
$S(z)$ is the flux of a source when placed at a redshift $z$.
The best-fit SED for the stellar mass was used to derive $S(z)$.
The comoving volume element $\frac{d^2V}{d\Omega dz}$
is calculated with formulae in \citet{CPT92} and \citet{H99}.
The redshift integration range is
determined by the criteria of the primary sample. 
$z_\mathrm{max}$ is the smaller of (a)
the redshift at which the source
has the minimum signal-to-noise ratio in the primary sample,
and (b) the upper end of the redshift bin.
$z_\mathrm{min}$ is the larger of (a)
the redshift at which the source
exceeds the maximum flux in the primary sample,
and (b) the lower end of the redshift bin.

We computed stellar mass functions
both for the spectroscopic sample
and for the combined sample of galaxies
with spectroscopic and photometric redshifts
(Table~\ref{tab:SM}).
Stellar mass functions for the spectroscopic sample
are listed to indicate the strict lower limits.
One sigma errors are also shown,
estimated from Poisson statistics
and uncertainties in $V_\mathrm{max}$.
The numbers in parentheses are
the number of galaxies that were used in the calculation.
Because of the small sample size,
we only used four broad stellar mass bins.
Some values are missing
due to the limited survey volume
(at the high stellar mass end at low redshift)
and limited sensitivity
(at the low stellar mass end at high redshift).
With such restrictions,
there are few overlaps among the redshift bins for a fixed stellar mass bin.
In each of such overlapping stellar mass bin,
we find that comoving space densities
are decreasing with redshift.
For a stellar mass range for typical local galaxies
($\log (M_\mathrm{star} [h_{65}^{-2}$\,M$_\odot])=10.95$--11.35),
the comoving space density at $z=1.2$--3.0
becomes 10\,\% ($-1.0$\,dex) of that at $z=0.2$--0.5.

\begin{figure*}
  \plotone{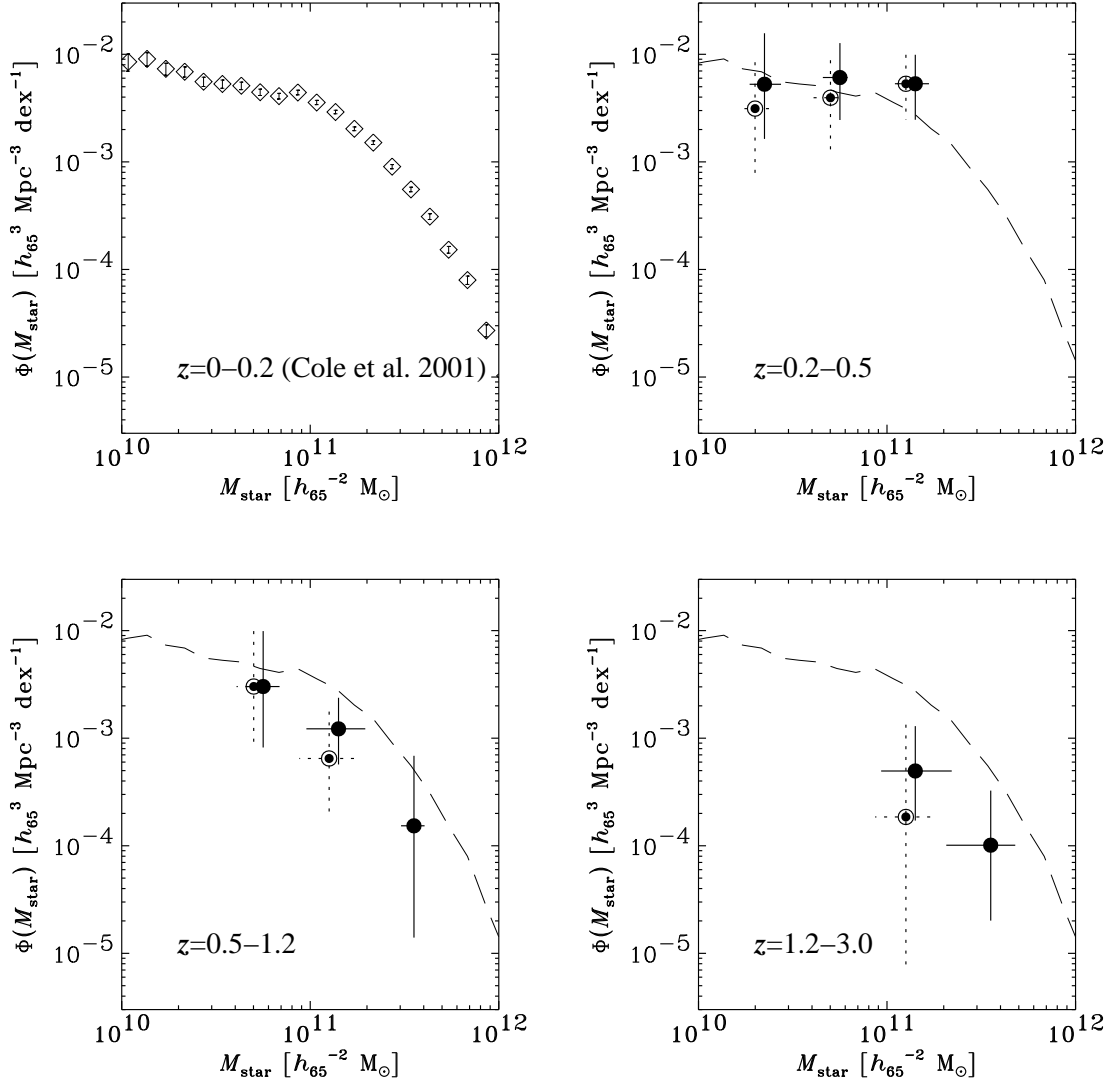}	
  \caption{
Stellar mass functions at four epochs.
For a local stellar mass function,
the stepwise maximum likelihood estimates for
a Salpeter IMF are reproduced from \citet{CNB+01}.
The accompanying errors are plotted,
though almost all are within symbols for values
(diamonds).
This local stellar mass function
is reproduced in three higher redshift panels (dashed lines).
In each of these panels,
our stellar mass function estimates
are shown with solid and double circles
for the combined (photometric and spectroscopic)
and for the spectroscopic samples, respectively.
Double circles are shifted slightly.
Vertical errors are
uncertainties in $V_\mathrm{max}$ and Poisson statistics.
Horizontal errors are mean fractional errors in stellar mass.
  }
\label{fig:SMF}
\end{figure*}

These stellar mass function estimates
are compared with local ones in Fig.~\ref{fig:SMF}.
\citet{CNB+01} derived a local stellar mass function
with a large sample of matched 2MASS-2dFGRS galaxies.
Their stepwise maximum likelihood estimates
are shown with diamonds in the upper-left panel.
The assumed initial mass function (IMF)
was a Salpeter-type
\citep{S55}.
The error bars are also shown.
The minimum error is obtained
almost at a characteristic stellar mass of
$M_\mathrm{star}^*=1.7 \times 10^{11}$\,M$_\odot$,
which is converted to our cosmology.
The redshift distribution of the 2MASS-2dFGRS galaxies is
largely confined to $z=0$--0.2 with a peak around $z=0.05$.
This local stellar mass function is represented with dashed lines
in higher redshift panels for the comparison with our estimates.
Solid and double circles show our stellar mass functions
derived with the combined and spectroscopic samples, respectively.
The values for the spectroscopic sample
are shifted leftward by 0.05\,dex.
Vertical error bars show Poisson noise.
Horizontal error bars are mean fractional errors in stellar masses.

In the $z=0.2$--0.5 bin,
our stellar mass function estimates are consistent with the local ones
within one sigma.
At higher redshifts,
stellar mass functions estimated with 6.7\,$\mu$m galaxies
start to show lower values than the local ones.
At redshifts of $z=0.5$--1.2,
the deviations are somewhat marginal.
The deviations appear to be larger at larger stellar masses.
At the highest redshift bin $z=1.2$--3.0,
the deviations from the local sample are more than one sigma.
These deviations could be much larger,
if we took account of our tendency to overestimate photometric redshifts
(Sect.~\ref{sect:zph}).
Cosmological surface brightness dimming could result in
a smaller number of galaxies
at high redshifts.
However, our sample has been selected from a 6.7\,$\mu$m map with a broad beam,
corresponding to a very deep surface brightness sensitivity
(0.02\,$\mu$Jy\,arcsec$^{-2}$ at $1\,\sigma$).

\begin{figure*}
  \plotone{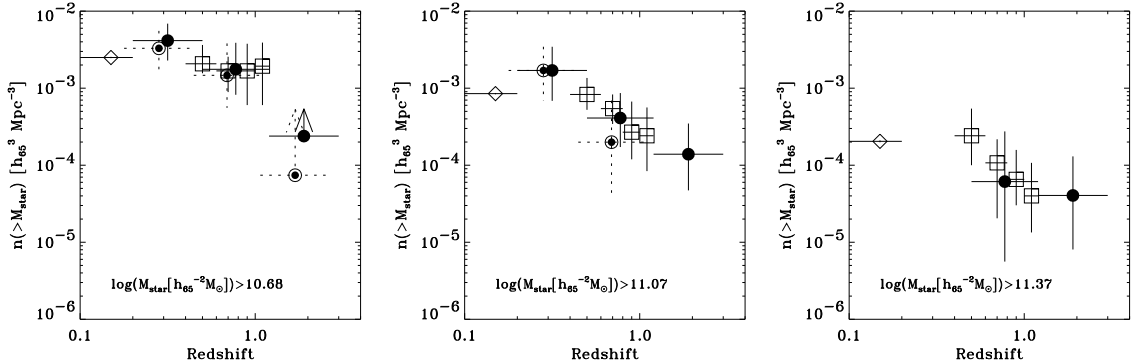}	
  \caption{
Redshift dependence of integrated stellar mass functions.
Lower stellar mass thresholds are those in \citet{DBS+01}.
Our estimates are shown as in Fig.~\ref{fig:SMF}.
\citet{DBS+01}'s values are represented as squares
with vertical error bars
indicating their field-to-field fluctuations.
Diamonds show local values
calculated from a stellar mass function by \citet{CNB+01}.
  }
\label{fig:ISMF}
\end{figure*}

The decrease in the comoving space density of massive stellar systems
at high redshifts has also been seen in $K$ band selected data
by \citet{DBS+01}.
They presented integrated stellar mass functions
above three mass thresholds
(Fig.~\ref{fig:ISMF}), which are almost comparable to
our largest three mass bins (Table~\ref{tab:SM}).
Their integrated stellar mass functions
(squares)
for $\log (M_\mathrm{star} [h_{65}^{-2}$\,M$_\odot])> 11.07$
and $\log (M_\mathrm{star} [h_{65}^{-2}$\,M$_\odot])> 11.37$
show declines of $-0.5$\,dex and $-0.8$\,dex
from $z=0.5$ to $z=1.1$.
Our integrated stellar mass functions
(circles)
have a good consistency in the overlapping redshift range.
For the lowest mass threshold panel,
our highest redshift bin values are lower limits,
because of the limited sensitivity
(Table~\ref{tab:SM}).
Note that \citet{DBS+01} adopted the maximum mass-to-light ratios
assuming the age of the universe.
The local values using \citet{CNB+01}'s stellar mass function
are marked as references
(diamonds).
Our values at $z=0.3$ are somewhat larger than \citet{CNB+01}'s values.
This difference may result from cosmic variance,
because our surveyed volume at such low redshift
is much smaller than that of \citet{CNB+01}.

\subsection{Stellar Mass Density in the Universe}

Using the stellar mass estimates for the 6.7\,$\mu$m galaxies,
we derived their contributions to the stellar mass density in the universe.
Epoch-dependent stellar mass densities $\rho_\mathrm{star} (z)$
are obtained as
\begin{equation}
\rho_\mathrm{star} (z) = \sum \frac{M_\mathrm{star}}{V_\mathrm{max}},
\end{equation}
utilizing the $V_\mathrm{max}$ values calculated in Sect.~\ref{sect:SMF}.
Here again, we only used the primary sample for the summation.
Our derived stellar mass densities for the three redshift bins
are direct sums.
Our stellar mass function estimates
have narrow mass ranges and large errors
(Fig.~\ref{fig:SMF}).
They could not be used to derive characteristic masses
and low-mass slopes,
which are needed to properly integrate stellar mass functions
over the full mass range.
Thus, we decided to use direct sums
in order to avoid large uncertainties in corrections to fully integrated values.
Our directly summed stellar mass densities
are shown in Table.~\ref{tab:SM}.
Their errors are estimated from
Poisson noise and uncertainties in stellar mass and $V_\mathrm{max}$.
The numbers of sources used in each summation are indicated in parentheses.
Stellar mass density becomes smaller at higher redshifts.
It should be noted that
the 6.7\,$\mu$m galaxies that contribute these values
come from different stellar mass ranges, as indicated in the left columns.

\begin{figure*}
  \plotone{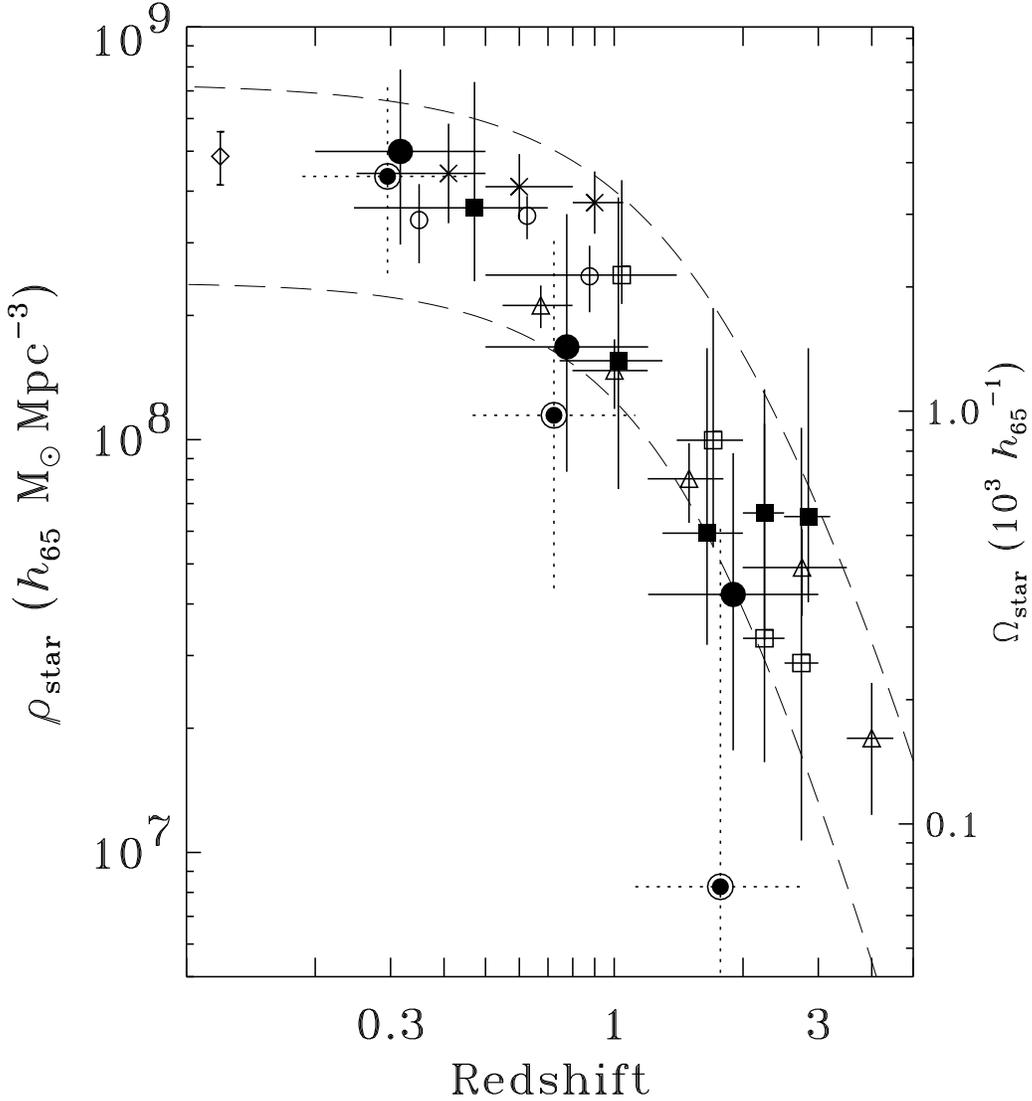}	
  \caption{
Stellar mass density in the universe as a function of redshift.
The right axis shows densities normalized to the critical density of the universe.
The contributions of the 6.7\,$\mu$m galaxies
are shown with solid and double circles
for the combined and spectroscopic samples, respectively.
The double circles are plotted at slightly lower redshifts.
The horizontal bars represent the redshift ranges of the bins
and the vertical bars show one sigma errors,
taking account of
Poisson noise and uncertainties in stellar mass and $V_\mathrm{max}$.
Several other estimates are overlaid;
solid squares are a $K$ band selected sample with photometric redshifts
\citep{FDV+03},
empty squares are an $H$ band selected sample with photometric redshifts
\citep{DPFB03},
the diamond is a $J$ band selected spectroscopic sample
\citep{CNB+01},
empty circles are an $I$ band selected spectroscopic sample
\citep{BE00},
X marks are an $R$ band selected spectroscopic sample
\citep{C02},
and
triangles are an $R^{\prime}$ band selected photometric sample
around a $z=4.7$ quasar
\citep{GDF+98}.
The empty/solid squares and the diamond are obtained
from full integration of a Schechter fit
to their respective luminosity or stellar mass function
at each redshift bin.
The X marks and the empty circles are quasi-fully integrated values
with a finite integration range from 10\,$L^*$ to 1/20\,$L^*$,
and $10.5<\log (M_\mathrm{star} [h_{65}^{-2}$\,M$_\odot])<11.6$,
respectively.
The triangles are simply summed values of the detected sources like ours.
We adopted her mean redshifts for \citet{C02}
and all four cases discussed in \citet{DPFB03} for their error bars.
Miller-Scalo IMFs were used in \citet{GDF+98}.
All points plotted here were normalized to our cosmology and
a Salpeter IMF with mass limits of 0.1\,M$_\odot$ and 125\,M$_\odot$.
The two dashed curves are deduced
by integrating the star formation rate density in the universe,
which is derived from the UV luminosity density as a function of redshift
\citep{CNB+01}.
The upper curve is an extinction corrected case for $E(B-V)=0.15$, and the lower one has no dust correction.
  }
\label{fig:SMD}
\end{figure*}

The contributions of our 6.7\,$\mu$m galaxies
to the stellar mass density in the universe
are shown as a function of redshift in Fig.~\ref{fig:SMD}.
Solid and double circles were estimated
from the combined and spectroscopic samples, respectively.
Double circles are shifted slightly to lower redshifts.
Horizontal bars show bin widths
and vertical bars mark one sigma uncertainties.
We overlay several stellar mass densities in the literature, as
triangles \citep{GDF+98},
empty circles \citep{BE00},
a diamond \citep{CNB+01},
X marks \citep{C02},
empty squares \citep{DPFB03},
and solid squares \citep{FDV+03}.

\begin{figure*}
  \plotone{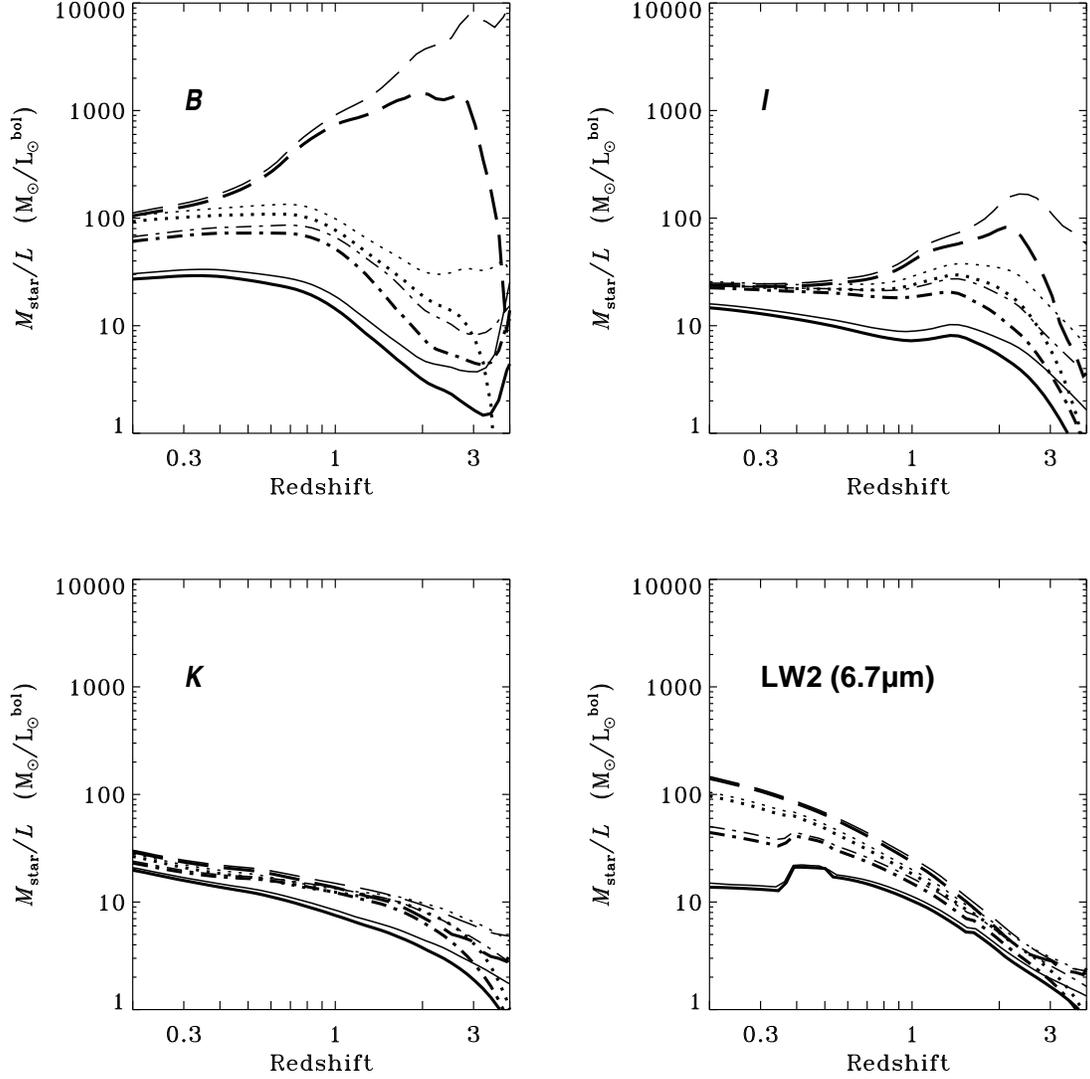}	
  \caption{
Stellar mass-to-light ratios as a function of redshift.
They are estimated for observations at
the $B$, $I$, $K$, and ISOCAM LW2 (6.7\,$\mu$m) bands.
The in-band luminosity at each observing band
corresponds to different wavelength with increasing redshift.
Thus, for the calculation of the ratios,
we adopt luminosities
in units of the bolometric luminosity of the Sun
(1\,L$_\odot^\mathrm{bol}=3.85 \times 10^{26}$\,[W]).
The GRASIL models for the evolving E, Sa, Sb, and Sc
galaxies are shown with
dashed, dotted, dot-dashed, and solid lines, respectively.
Two formation redshifts are assumed;
$z_\mathrm{f}=5$ and 10
(thick and thin lines, respectively).
  }
\label{fig:z_ml4}
\end{figure*}

The local value by \citet{CNB+01}
was obtained from their stellar mass function,
which is deduced assuming a Salpeter IMF.
The value is derived for the full mass range
by integrating the Schechter fit to the stellar mass function.
\citet{DPFB03} also adopted fully integrated values
over the full mass range.
They first derived luminosity densities
by integrating the Schechter fit
to their rest-frame $B$-band luminosity function at each redshift,
and then converted them with their mean $B$-band mass-to-light ratio
into stellar mass densities.
\citet{FDV+03} followed this method.
\citet{C02} used the integration
of the Schechter fits to the luminosity functions,
but for a restricted range from 10\,$L^*$ to 1/20\,$L^*$.
These quasi-integrated values for her $K$-band luminosity functions
were converted with a fixed $z=0$ stellar mass-to-light ratio of
$M_\mathrm{star}/L_K$=0.8 in solar units.
Note that this $M_\mathrm{star}/L_K$ ratio has a different meaning
from our $M_\mathrm{star}/L$ ratio for the $K$ band
in units normalized to the bolometric solar luminosity
(Sect.~\ref{sect:SMLR}),
which is shown in Fig.~\ref{fig:z_ml4}.
\citet{BE00} also adopted the value of $M_\mathrm{star}/L_K$=0.8
to derive incompleteness corrections.
Their values are quasi-full integrations
for a limited mass range of
$10.5<\log (M_\mathrm{star} [h_{65}^{-2}$\,M$_\odot])<11.6$
(cf. Table~\ref{tab:SM}).
\citet{GDF+98}'s values are direct sums
for $20<R^{\prime}<25$ galaxies around a $z=4.7$ quasar.
Their values are multiplied by two
to take account of their adopted Miller-Scalo IMF.

Some authors assumed a Schechter form;
however, the shape of stellar mass functions is not determined yet.
We therefore did not applied any corrections
to these values derived for the different mass ranges.
We made crude corrections only for the different cosmologies.
Although the reported values are basically binned values,
we took each of them as a single representative value
at each redshift with a fixed redshift range.
We did not consider effects from sources near the boundaries of the bins
or with different $z_\mathrm{min}$ and $z_\mathrm{max}$ values.
For \citet{BE00},
we used midpoints in $\log z$ for their bin boundaries.
We used her mean $z$ values for \citet{C02}.

At lower redshifts ($z<0.5$),
almost all the estimates are consistent
and very near to the local value.
This suggests that
surveys cited here have succeeded in detecting
most of the stellar masses in galaxies at these redshifts.
At a higher redshift ($z \sim 1$),
differences between the results become somewhat larger.
The use of a fixed stellar mass-to-ratio of $M_\mathrm{star}/L_K=0.8$
in \citet{C02} and \citet{BE00}
could explain some of the values in excess of ours at $z \sim 1$.
According to \citet{DBS+01},
stellar mass-to-ratio $M_\mathrm{star}/L_K$
is a decreasing function of redshift;
0.99, 0.88, 0.73, and 0.65 for $z=0.5$, 0.7, 0.9, and 1.1, respectively.
They assumed the maximum ages for the passively evolving galaxies;
thus, the decline in $M_\mathrm{star}/L_K$ could be steeper.
Note that the declines in the $M_\mathrm{star}/L$ ratios at the $K$ band
in Fig.~\ref{fig:z_ml4} come from the evolution of the galaxies
but also from the effects of the shifting observing band.
Direct sums of the detected sources only
are likely to miss low mass galaxies below the detection limits.
Therefore, our spectroscopic points,
which are derived from our sub-samples,
should be regarded as strict lower limits.
At even higher redshifts ($z>2$),
\citet{GDF+98}'s values appear to have a milder decline,
especially compared with those of \citet{DPFB03}.
This might related to their use of the different IMF in the estimation.
However, \citet{FDV+03} presented
high stellar mass densities at these redshifts.
This suggests that we need to survey much larger area to determine
the true stellar mass densities at high redshift.

The evolution of stellar mass density
will be related to that of the star formation rate density in the universe.
Star formation rate indicators at the UV wavelengths
are sensitive to dust extinction.
\citet{CNB+01} adopted two cases; $E(B-V)=0$ and $E(B-V)=0.15$.
For each case,
they provided an analytic formula
fitted to the UV observations of the star formation rate density.
By integrating these formulae with time,
we estimated the evolution of stellar mass density in the universe.
Here, the recycling fraction of stellar mass
for the next generation stars was assumed to be $R=0.28$
for a Salpeter IMF.
The results are overlaid with dashed lines in Fig.\ref{fig:SMD}.
Lower and upper lines correspond to
the $E(B-V)=0$ and $E(B-V)=0.15$ cases, respectively.
It should be noted that these lines
are shown with units of M$_\odot\,\mathrm{Mpc}^{-3}$
with no dependence on the Hubble parameter
because of the cancellation in the time integration.

These time-integrated values of the star formation rate densities
are derived by the integration of
the full range of star formation rate or luminosity at each epoch.
When we assume $E(B-V)=0.15$,
the line calculated from the time integration of
the star formation rate densities
becomes always higher than
any of the stellar mass density points
estimated for the full or the quasi-full range of stellar mass.
This might suggest that the mean dust extinction value
would be lower than $E(B-V)=0.15$.
In fact, a median $E(B-V)$ value for the $z \sim 4$ Lyman break galaxies
is somewhat lower than $E(B-V)=0.15$
\citep{SAG+99}.
However, it should be noted
that all the full or the quasi-full integrated values
for the stellar mass densities at high redshifts
were derived from the rest-frame UV or optical light only.
In addition to 
underestimation of this light due to a certain level of
dust extinction,
there remains some possibility that
highly reddened systems were completely neglected due to their non-detections.
In fact, a star formation rate density estimate
comparable to the values for the case of $E(B-V)=0.15$
was obtained with only a few submillimeter sources
\citep{HSD+98}.
No contributions from such dusty galaxies were taken into account
in the \citet{CNB+01} formulae.
It should be noted that
our stellar mass density estimates
include contributions from the three submillimeter galaxies in this field.
At 6.7\,$\mu$m,
we can probe the rest-frame near-infrared light out to quite a high redshift.
Because dust extinction at the rest-frame near-infrared is almost negligible,
underestimates of stellar mass densities are unlikely to happen
either by loss of light or by non-detections.

\section{DISCUSSION}
\label{sect:dis}

The contributions of the faint 6.7\,$\mu$m galaxies
to the stellar mass density in the universe
are estimated to be comparable to
those inferred from observations of UV bright galaxies.
Unfortunately,
6.7\,$\mu$m observations probe a narrow mass range
and UV observations suffer from dust effects,
preventing us from making detailed comparisons between the two.

On the other hand,
we found that the faint 6.7\,$\mu$m galaxies
generally had red colors.
A comparison with a particular population synthesis model
suggests that they have experienced
vigorous star formation at high redshifts.
The derived large stellar masses
for the faint 6.7\,$\mu$m galaxies
also support such star forming events at the past.
Beyond the redshift range of our sample ($z>3$),
we know that there exist Lyman break galaxies.
They are blue and forming stars;
however, their masses are generally
smaller than the masses of this faint 6.7\,$\mu$m sample.
In a naive sense,
several Lyman break galaxies
must merge to form a massive 6.7\,$\mu$m galaxy.
Or, very rapid star forming systems are needed.
SCUBA galaxies are expected to have
such efficient star forming activities.
However, at least in this field,
SCUBA sources were already identified as faint 6.7\,$\mu$m galaxies
at relatively small redshifts.

We noticed the existence of massive galaxies out to $z=3$.
But at the same time,
their comoving space densities were found to be lower
than the present values.
Thus, there should be some mass assembly activities
to build up massive galaxies in a redshift range of $z=0$--3.
They will be forming stars in situ, merging, or
a combination of the two.
The investigation of such build-ups in $z=0$--3
would give us an important insights
to the build-ups in the higher redshift regime.
A large number of star forming galaxies detected at 15\,$\mu$m with ISO
may give an unbiased sample for this purpose
because of their insensitivity to dust.

The detection of all the X-ray, submillimeter, and radio
sources in this field at 6.7\,$\mu$m is interesting.
Most of our knowledge of the evolution of galaxies
has been based on investigation of stellar systems,
such as optical observations of UV emission from
massive stars.
But a higher fraction of active galaxies detectable at other wavelengths
in the distant universe
requires the consistent understanding of the evolution of multiple
components in galaxies.
Far-infrared/submillimeter observations of dust
and X-ray/radio observations of active galactic nuclei (AGN)
might provide us with more essential information
than UV/optical/near-infrared observations for stellar components.

We have stated that
mid-infrared observations resulted in
an investigation of a narrow range of stellar masses.
We regard this as a good property,
i.e., allowing the preparation of a mass-ordered sample
to compare stars/dust/AGN with multi-wavelength observations.
With mid-infrared surveys in the near future,
we expect to construct well-controlled samples of distant galaxies.

\section{CONCLUSIONS}
\label{sect:con}

The tight correlation
between stellar masses of galaxies
and their rest-frame near-infrared luminosities
can be a strong tool for investigating
the evolution of stellar mass assembly in galaxies.
The effect of redshift has motivated us
to observe high redshift galaxies in the mid-infrared.
For the mid-infrared sources detected in the SSA13 field
with the ISOCAM LW2 (6.7\,$\mu$m) filter,
their nature and stellar masses were discussed
after their identifications.
The 65 sources were divided into two subsamples,
a primary sample of 33 sources (of which 2 are stars)
and a supplementary sample of 32 sources.
No spurious sources are expected in the primary sample.
Taking account of
higher source densities at the optical and near-infrared,
the identifications at these wavelengths
were determined by utilizing two possibilities;
one for the true association and the other for a chance event.
Using the highest ratio of the two,
32 out of the 33 primary sources
and 25 out of the 32 supplementary sources were identified.
A test of the identification procedure with the negative sample
showed that for the supplementary sample
identifications at the $K$ band should be secure,
while a few identifications at the $I$ band might be wrong.
A comparison with
the published X-ray, submillimeter, and radio catalogues
resulted in mid-infrared identifications of
all (four) X-ray, (three) submillimeter, and (one) radio sources in the field.
They were all in the primary sample.

With the color information from the optical and near-infrared identifications,
we can divide 6.7\,$\mu$m galaxies into three types.
The GRASIL galaxy model predicts that
red $B-I$ colors and blue $K-6.7\,\mu\mathrm{m}$ colors
can be used to isolate low redshift early type galaxies
(type~I).
Red $B-K$ colors and red $I-6.7\,\mu\mathrm{m}$ colors
were used to select high redshift early type progenitors (type~II),
which can be regarded as ancestors of the type~Is.
Blue $B-K$ colors and blue $I-6.7\,\mu\mathrm{m}$ colors
are an indicator of on-going star formation (type~III).
The main contributors to the type~III category would be late type galaxies.
The ratio of the three type was almost $1:1:1$ in the 31 primary galaxies,
while the supplementary galaxies have a higher fraction of type~II.

In order to permit quantitative discussions,
we estimated photometric redshifts for the 6.7\,$\mu$m galaxies.
Based on their good representations in the color-color plots,
we used a limited set of SEDs
in the GRASIL library
for the fitting templates.
A test of this photometric redshift estimation
with a spectroscopic subsample
suggested that they could be used as good redshift estimates.
Although a few of them could be high redshift outliers,
the large errors in their photometric redshifts
might be used as indicators of such.

Modulo the caveats on photometric redshifts,
we found that deep 6.7\,$\mu$m surveys
were efficient in detecting high redshift galaxies.
The flux redshift relation of the primary 6.7\,$\mu$m galaxies
showed a high redshift tail at fluxes below 30\,$\mu$Jy.
A significant fraction of the photometric redshifts
for the supplementary galaxies were $z>1$,
consistent with the dominance of type~II in them.
A $K<20$ sample has low redshift galaxies with $K>18$ and $I-K<2$,
while our 6.7\,$\mu$m sample
do not include such a population of blue low-mass galaxies.

Stellar masses were derived utilizing a tight correlation
between rest-frame near-infrared luminosity and stellar mass.
Stellar mass-to-light ratios were determined
from fits to a limited set of template SEDs
in the GRASIL library.
With a hybrid conversion to stellar mass,
using $K$ luminosities at $z<1$
and 6.7\,$\mu$m luminosities at $z>1$,
we estimated stellar masses for the 6.7\,$\mu$m galaxies.
We found that some of the high redshift 6.7\,$\mu$m galaxies
had stellar masses comparable to
the typical stellar mass of local galaxies.
However, the comoving space density of such massive galaxies
is likely to be a decreasing function of redshift.
The epoch-dependent stellar mass functions
might suggest that more massive galaxies are rarer at higher redshifts.
If our photometric redshifts are slight overestimates,
this trend will be even stronger.

We derived the contributions of the 6.7\,$\mu$m galaxies
to the stellar mass density in the universe as a function of redshift.
Given the narrow mass ranges,
our estimates were obtained as simple summations of the detected sources.
Our low redshift value was almost consistent with the local value,
suggesting that our sample includes major contributors
to the low redshift stellar mass density.
At the same time,
most of the mass assembly in galaxies should be finalized at this epoch
($z \sim 0.3$).
The stellar mass density estimates become smaller at higher redshifts.
This would be expected from
the decrease in the high redshift stellar mass functions;
however, the density we compute for the highest redshift bin
is almost comparable to
the full mass range value for a rest-frame optical selected sample.
Note that our value includes contributions of dusty submillimeter galaxies.
A full mass range sample of rest-frame near-infrared selected galaxies
will be necessary to estimate correct values of
the stellar mass density
taking into account any dusty population.

\acknowledgments

We are indebted to an anonymous referee for critical
but kind and thorough comments.
YS would like to thank Toru Yamada, Joel Primack,
Peter Eisenhardt, Mark Dickinson, and Alberto Franceschini
for their encouragement at the ESO/Venice workshop
on the mass of galaxies in 2001.
He also acknowledges Masayuki Uemura and Nobuo Arimoto
for their comments at the initial phases of this work.
This research has been partly supported by
JSPS Research Fellowships for Young Scientists.
The analysis has been achieved with the IDL Astronomy Users Library
maintained by Wayne Landsman.
This research has
made use of NASA's Astrophysics Data System Bibliographic Services.

\appendix

\section{THE GRASIL SED LIBRARY}
\label{sect:GRASIL}

Granato and Silva introduced effects of Graphite and Silicate dust
to a SED model of galaxies
based on the stellar population synthesis method
\citep{SGBD98}.
Effects of dusty interstellar media are treated with
a radiative transfer code.
Thus, the GRASIL SEDs cannot be arbitrarily rescaled.
The modeled SEDs include UIB emissions,
which are believed to be caused by
policyclic aromatic hydrocarbon (PAH) particles.
A small suite of SEDs and executables to calculate SEDs
are publicly available
at the official GRASIL web site
(\texttt{http://web.pd.astro.it/granato/grasil/grasil.html}).
The full details of the model can be traced at that site
and the SED library itself is expanding.
In order to avoid generating inadequate SEDs
by using executables with excessive parameter sets,
we just have used the author-proofed SED library in this paper.
Here we describe the salient properties of
two sets of SEDs in the library.

\begin{figure*}
  \plotone{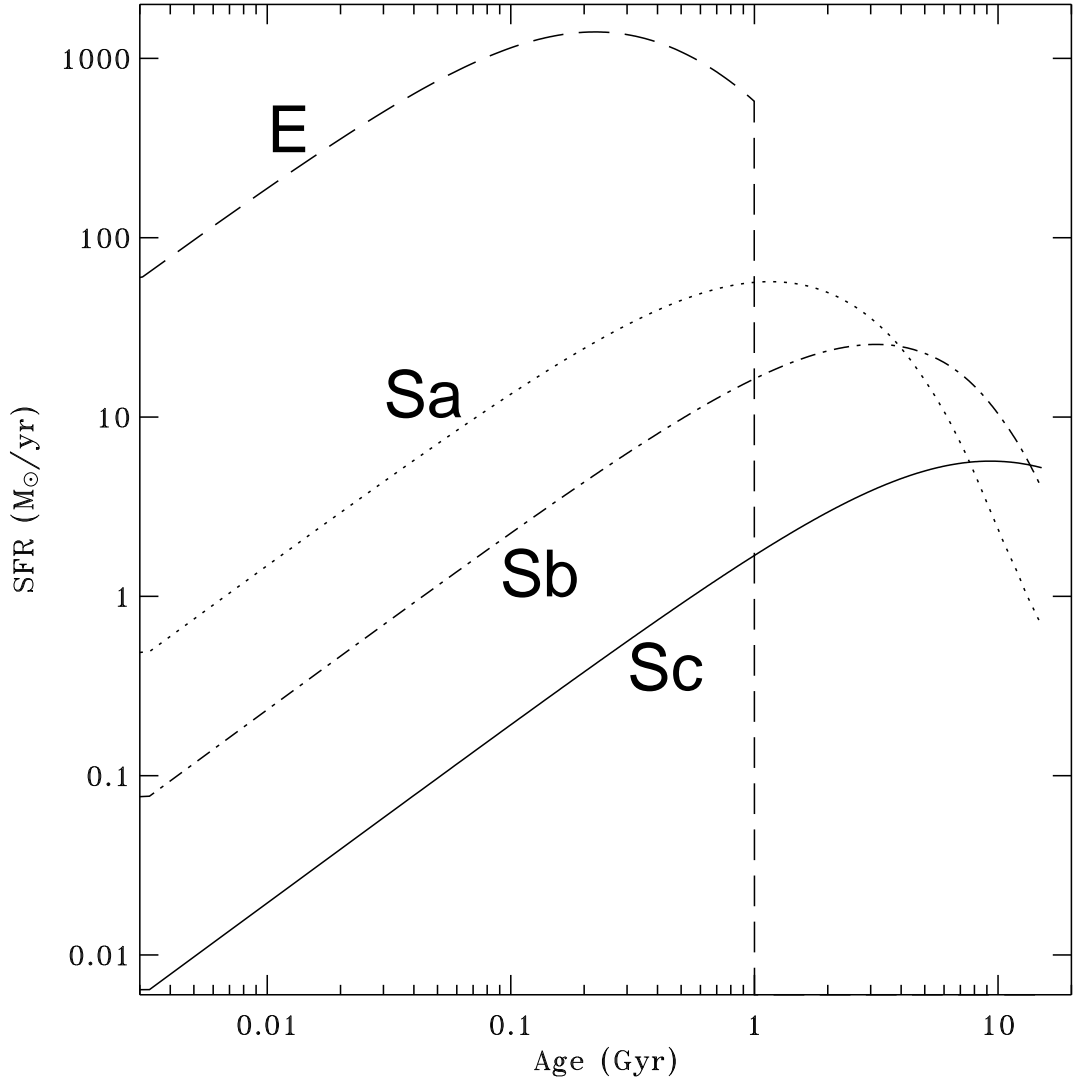}	
  \caption{
Star formation rates (SFRs) for the evolving E, Sa, Sb, and Sc galaxies
in the GRASIL library
(Appendix~\ref{sect:GRASIL}).
They are shown as a function of age
with dashed, dotted, dot-dashed, and solid lines, respectively.
It should be noted that age increases from left to right.
After the age of 1\,Gyr, the E galaxy was assumed to stop its star formation
because of a galactic wind.
  }
\label{fig:sfr}
\end{figure*}

One is a set of evolving SEDs for four Hubble types
in the local universe; E, Sa, Sb, and Sc.
Elliptical galaxies are assumed to be formed
in a monolithic collapse scenario.
The classification of spiral galaxies is based on \citet{SSS89}.
These SEDs span from UV to radio wavelengths.
The number of ages are 12 for the E galaxy
(0.1, 0.2, 0.4, 0.8, 1.5, 2, 3, 4, 5, 8, 11, 13\,Gyr)
and 15 for the Sa, Sb, and Sc galaxies
(1, 2, 3, 4, 5, 6, 7, 8, 9, 10, 11, 12, 13, 14, 15\,Gyr).
Their star formation histories are assumed to be smooth,
as shown in Fig.~\ref{fig:sfr}.
By setting a cosmology and a formation redshift,
we can convert their age into redshift.

For stellar populations,
Salpeter IMFs \citep{S55} are assumed
with mass limits of 0.15\,M$_\odot$ and 120\,M$_\odot$ for the E galaxy
and 0.10\,M$_\odot$ and 100\,M$_\odot$ for the Sa, Sb, and Sc galaxies.
Using a $K$ band luminosity function for local galaxies,
\citet{CNB+01} have provided a stellar mass function
for a Salpeter IMF with mass limits of 0.1\,M$_\odot$ and 125\,M$_\odot$.
The characteristic stellar mass for this stellar mass function
becomes $M_\mathrm{star}^*=1.7 \times 10^{11}$\,M$_\odot$
for our adopted cosmology.
With this unit,
the GRASIL E, Sa, Sb, and Sc galaxies
with a formation redshift of $z_\mathrm{f}=2$
have stellar masses of
$\sim 3\,M_\mathrm{star}^*$,
$\sim 0.8\,M_\mathrm{star}^*$,
$\sim 0.7\,M_\mathrm{star}^*$, and $\sim 0.2\,M_\mathrm{star}^*$
at $z=0$, respectively.
All the stellar mass estimates with the GRASIL SEDs in this paper
were normalized to
a Salpeter IMF with mass limits of 0.1\,M$_\odot$ and 125\,M$_\odot$.

The other set of SEDs contains fitted templates of nearby galaxies,
which are detailed in \citet{SGBD98}.
The model parameters were chosen to represent SEDs for
starburst galaxies (M82, NGC6090, and Arp220),
normal galaxies (M51, M100, and NGC6946),
and a giant elliptical.
The wavelength range is shorter than the evolving SEDs above,
from UV to submillimeter.

We used all the SEDs above to derive photometric redshifts.
To obtain stellar masses, we only utilized the evolving SEDs,
since some of the fitted templates lack stellar mass information.
Model predictions based on the evolving SEDs are shown in some plots,
though the Sb SED was omitted to avoid overcrowding.

\clearpage

\begin{figure*}								
  \plotone{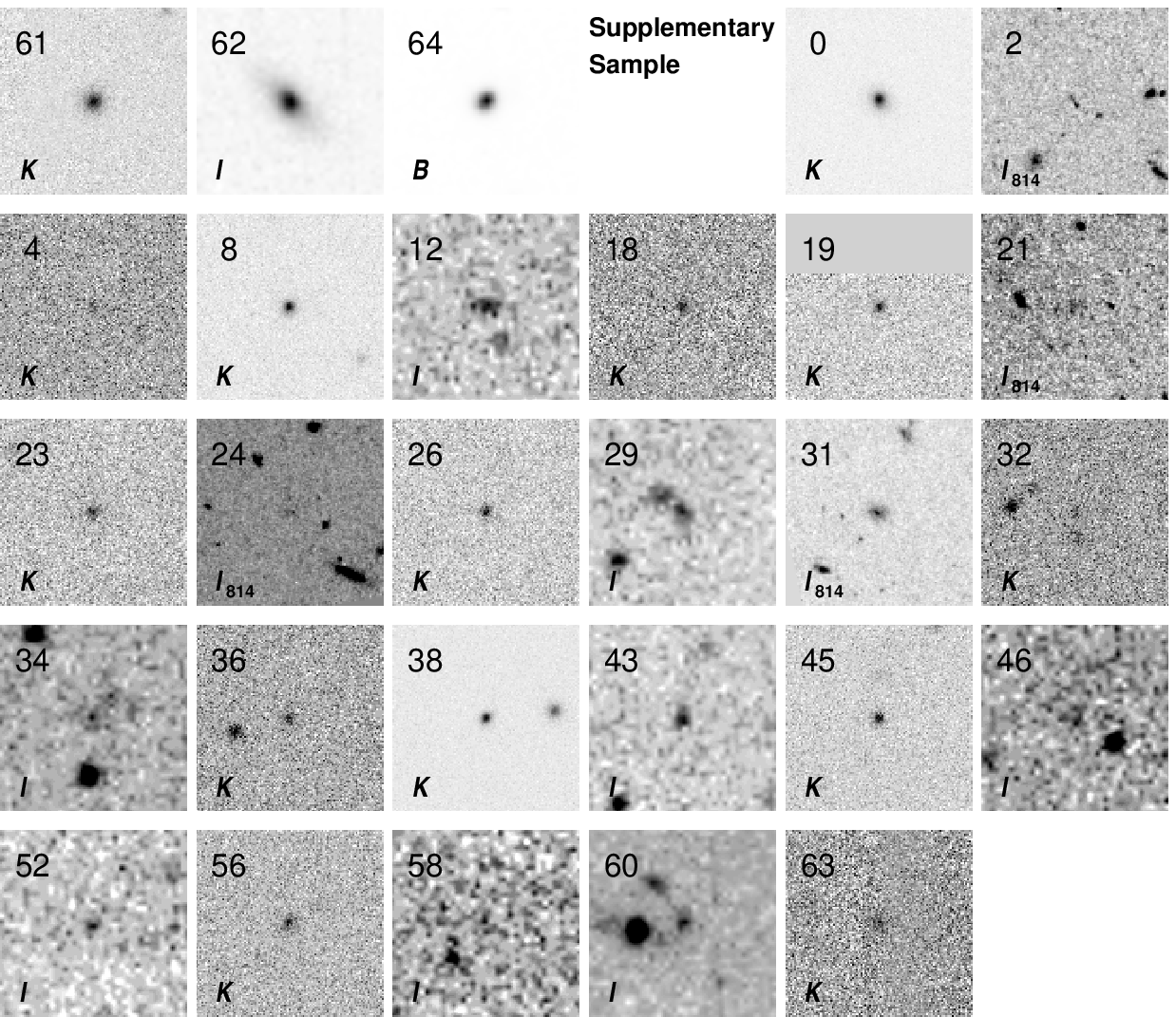}							
  \vspace{1mm}\leftline{Fig. \ref{fig:id}.--- Continued.}\vspace{4mm}	
\end{figure*}								

\clearpage

\LongTables				

\clearpage
\def\arraystretch{1.12}			
\tabletypesize{\scriptsize}		
\begin{deluxetable*}{rcccrrlcllll}	
\tablecolumns{10}
\tablewidth{0pt}
\tablecaption{
Source Identifications in the Optical and Near-Infrared
  \label{tab:id}
  }
\tablehead{
\colhead{Name}			&
\colhead{R.A.}			& \colhead{Decl.}		& \colhead{$r_0$}		&
\colhead{6.7\,$\mu$m}		& \colhead{S/N}			&
\colhead{ID}			& \colhead{$r_1$}		&
\colhead{$P_\mathrm{t}$}	& \colhead{$P_\mathrm{c}$}	&
\colhead{$P_\mathrm{t}/P_\mathrm{c}$}	& \colhead{$P_\mathrm{t}/P_\mathrm{c}^{(2)}$}	\\
\colhead{\#}			&
\multicolumn{2}{c}{(J2000.0)}					& \colhead{(\arcsec)}		& 
\colhead{($\mu$Jy)}		& \colhead{}			&
\colhead{}			& \colhead{(\arcsec)}		&
\colhead{}			& \colhead{}			&
\colhead{}			& \colhead{}
}
\startdata
 & & & & & & & & & & & \\
\sidehead{Primary Sample}
 3 & 13 12 18.09 & +42 43 45.0 & 1.1 & $  50$ & $ 11.4$ & $K$       & 0.8 &  0.49 & 0.000056 & \phm{$00$}880. &         \\
 6 & 13 12 18.32 & +42 43 19.2 & 1.1 & $  66$ & $ 11.9$ & $K$       & 0.5 &  0.63 & 0.000038 & \phm{$0$}1700. &         \\
 9 & 13 12 19.48 & +42 45 36.4 & 1.5 & $  25$ & $  5.8$ & $K$       & 0.5 &  0.75 & 0.00069 & \phm{$00$}110. &         \\
10 & 13 12 20.01 & +42 44 38.4 & 1.4 & $  21$ & $  6.4$ & $K$       & 0.7 &  0.64 &  0.0015 & \phm{$000$}41. &         \\
11 & 13 12 21.02 & +42 44 33.8 & 1.5 & $  19$ & $  5.9$ & $K$       & 1.0 &  0.49 & 0.00094 & \phm{$000$}53. & \phm{$0$}3.7 \\
13 & 13 12 21.39 & +42 44 23.4 & 1.4 & $  24$ & $  7.0$ & $K$       & 0.3 &  0.80 & 0.000092 & \phm{$00$}870. &         \\
14 & 13 12 21.57 & +42 44 05.8 & 1.5 & $  19$ & $  6.3$ & $K$       & 0.8 &  0.58 &  0.0023 & \phm{$000$}25. &         \\
15 & 13 12 21.58 & +42 45 18.8 & 1.0 & $  60$ & $ 15.6$ & $K$       & 0.3 &  0.77 & 0.000086 & \phm{$00$}900. &         \\
16 & 13 12 21.89 & +42 43 45.5 & 1.5 & $  19$ & $  5.8$ & $K$       & 2.3 &  0.12 &   0.031 & \phm{$0000$}3.8 &         \\
17 & 13 12 22.53 & +42 44 50.9 & 1.5 & $  17$ & $  5.3$ & $K$       & 0.3 &  0.86 & 0.00053 & \phm{$00$}160. & \phm{$0$}0.14 \\
20 & 13 12 23.65 & +42 45 16.9 & 1.5 & $  20$ & $  6.2$ & $K$       & 2.9 & 0.048 &   0.032 & \phm{$0000$}1.5 &         \\
22 & 13 12 23.91 & +42 45 43.5 & 0.9 & $ 135$ & $ 28.5$ & $K$       & 1.1 &  0.25 & 0.0000023 & 11000. &         \\
25 & 13 12 24.90 & +42 44 14.8 & 1.1 & $  39$ & $ 11.7$ & $K$       & 2.5 & 0.023 &  0.0014 & \phm{$000$}16. &         \\
27 & 13 12 25.2\phm{$0$} & +42 46 00\phm{$.0$} & 1.2 & $  45$ & $ 10.1$ & $ $       &     &       &                        &                       &         \\
28 & 13 12 25.18 & +42 43 44.9 & 1.3 & $  27$ & $  7.9$ & $K$       & 1.3 &  0.34 &  0.0042 & \phm{$000$}82. & \phm{$0$}2.8 \\
30 & 13 12 26.31 & +42 42 26.9 & 1.6 & $  70$ & $  5.0$ & $I$       & 0.9 &  0.59 & 0.000025 & \phm{$0$}2300. &         \\
33 & 13 12 27.32 & +42 44 49.7 & 1.5 & $  17$ & $  5.4$ & $K$       & 1.4 &  0.36 &  0.0087 & \phm{$000$}41. & \phm{$0$}0.032 \\
35 & 13 12 27.70 & +42 45 36.6 & 1.7 & $  15$ & $  4.8$ & $K$       & 1.3 &  0.44 &  0.0059 & \phm{$000$}75. &         \\
39 & 13 12 28.44 & +42 46 03.3 & 1.5 & $  27$ & $  5.4$ & $I$       & 1.4 &  0.35 &  0.0078 & \phm{$000$}45. &         \\
40 & 13 12 28.29 & +42 44 54.6 & 1.7 & $  14$ & $  4.8$ & $K$       & 3.4 & 0.047 &   0.084 & \phm{$0000$}0.56 &         \\
41 & 13 12 28.58 & +42 43 58.9 & 1.1 & $  46$ & $ 13.3$ & $K$       & 1.0 &  0.37 & 0.00016 & \phm{$00$}240. &         \\
44 & 13 12 29.21 & +42 45 58.2 & 1.3 & $  37$ & $  8.2$ & $I$       & 1.1 &  0.42 & 0.00032 & \phm{$00$}130. &         \\
47 & 13 12 29.88 & +42 44 08.5 & 1.5 & $  16$ & $  5.3$ & $K$       & 0.7 &  0.64 & 0.00049 & \phm{$00$}130. & \phm{$0$}0.19 \\
49 & 13 12 30.10 & +42 44 20.7 & 1.6 & $  15$ & $  5.0$ & $K$       & 1.3 &  0.42 &  0.0071 & \phm{$000$}60. & \phm{$0$}0.14 \\
50 & 13 12 30.16 & +42 44 56.5 & 1.5 & $  19$ & $  5.8$ & $K$       & 2.4 &  0.11 &   0.015 & \phm{$0000$}7.3 & \phm{$0$}0.11 \\
51 & 13 12 31.05 & +42 43 32.6 & 1.4 & $  22$ & $  6.7$ & $K$       & 0.6 &  0.66 & 0.00025 & \phm{$00$}260. &         \\
54 & 13 12 31.50 & +42 45 52.2 & 0.9 & $ 167$ & $ 23.6$ & $K$       & 1.2 &  0.20 & 0.000095 & \phm{$00$}210. &         \\
55 & 13 12 31.90 & +42 43 46.4 & 1.9 & $  12$ & $  4.3$ & $K$       & 2.1 &  0.25 &   0.010 & \phm{$000$}25. &         \\
57 & 13 12 31.94 & +42 44 29.7 & 1.5 & $  18$ & $  6.2$ & $K$       & 2.2 &  0.13 &   0.033 & \phm{$0000$}3.9 &         \\
59 & 13 12 33.92 & +42 44 42.7 & 1.6 & $  18$ & $  5.0$ & $K$       & 1.2 &  0.47 & 0.00099 & \phm{$000$}47. & 20. \\
61 & 13 12 34.68 & +42 43 42.8 & 1.3 & $  55$ & $  8.3$ & $K$       & 2.5 & 0.058 &  0.0024 & \phm{$000$}24. &         \\
62 & 13 12 34.51 & +42 43 09.3 & 1.8 & $  88$ & $  4.5$ & $I$       & 2.1 &  0.25 & 0.00056 & \phm{$000$}45. &         \\
64 & 13 12 35.90 & +42 43 58.0 & 1.3 & $ 107$ & $  7.7$ & $B$       & 2.0 &  0.14 & 0.000074 & \phm{$00$}190. &         \\
\sidehead{Supplementary Sample}
 0 & 13 12 15.79 & +42 44 00.3 & 2.5 & $  19$ & $  2.7$ & $K$       & 0.6 &  0.80 & 0.00013 & \phm{$00$}600. &         \\
 1 & 13 12 16.8\phm{$0$} & +42 44 21\phm{$.0$} & 1.9 & $  18$ & $  4.1$ & $ $       &     &       &                        &                       &         \\
 2 & 13 12 17.10 & +42 44 52.2 & 2.3 & $  14$ & $  3.0$ & $I_{814}$ & 3.1 &  0.18 &    0.11 & \phm{$0000$}1.6 & \phm{$0$}0.17 \\
 4 & 13 12 17.82 & +42 45 27.3 & 2.3 & $  18$ & $  3.2$ & $K$       & 2.6 &  0.25 &   0.044 & \phm{$0000$}5.6 &         \\
 5 & 13 12 18.1\phm{$0$} & +42 46 15\phm{$.0$} & 2.6 & $  34$ & $  2.7$ & $ $       &     &       &                        &                       &         \\
 7 & 13 12 18.7\phm{$0$} & +42 42 48\phm{$.0$} & 2.3 & $  66$ & $  3.0$ & $ $       &     &       &                        &                       &         \\
 8 & 13 12 19.34 & +42 45 01.1 & 1.9 & $  12$ & $  4.0$ & $K$       & 2.0 &  0.29 &  0.0065 & \phm{$000$}44. &         \\
12 & 13 12 21.19 & +42 46 26.0 & 2.3 & $  63$ & $  3.0$ & $I$       & 2.0 &  0.39 &   0.015 & \phm{$000$}26. & \phm{$0$}4.0 \\
18 & 13 12 22.45 & +42 43 15.6 & 2.7 & $   8$ & $  2.6$ & $K$       & 4.1 &  0.14 &   0.062 & \phm{$0000$}2.2 &         \\
19 & 13 12 23.10 & +42 46 05.4 & 2.2 & $  17$ & $  3.3$ & $K$       & 1.9 &  0.39 &  0.0067 & \phm{$000$}58. &         \\
21 & 13 12 23.30 & +42 44 29.4 & 2.3 & $   9$ & $  3.0$ & $I_{814}$ & 2.3 &  0.31 &   0.081 & \phm{$0000$}3.9 & \phm{$0$}2.1 \\
23 & 13 12 24.18 & +42 43 57.7 & 2.3 & $   9$ & $  3.1$ & $K$       & 2.0 &  0.39 &   0.022 & \phm{$000$}18. &         \\
24 & 13 12 24.33 & +42 43 23.9 & 2.0 & $  12$ & $  3.8$ & $I_{814}$ & 2.0 &  0.31 &   0.078 & \phm{$0000$}3.9 & \phm{$0$}0.067 \\
26 & 13 12 24.60 & +42 45 11.4 & 2.1 & $  10$ & $  3.5$ & $K$       & 2.2 &  0.29 &   0.023 & \phm{$000$}13. &         \\
29 & 13 12 25.56 & +42 43 20.5 & 2.4 & $   8$ & $  2.8$ & $I$       & 2.4 &  0.33 &   0.016 & \phm{$000$}20. & \phm{$0$}8.5 \\
31 & 13 12 26.58 & +42 45 30.0 & 3.2 & $   6$ & $  2.4$ & $I_{814}$ & 5.2 &  0.11 &   0.086 & \phm{$0000$}1.2 & \phm{$0$}0.43 \\
32 & 13 12 26.31 & +42 44 04.8 & 2.1 & $   9$ & $  3.4$ & $K$       & 1.3 &  0.55 &   0.011 & \phm{$000$}48. & \phm{$0$}1.5 \\
34 & 13 12 27.51 & +42 46 30.4 & 2.4 & $  45$ & $  2.9$ & $I$       & 2.0 &  0.40 &   0.017 & \phm{$000$}23. & \phm{$0$}4.1 \\
36 & 13 12 27.80 & +42 45 09.6 & 2.4 & $   8$ & $  2.8$ & $K$       & 1.0 &  0.66 &  0.0053 & \phm{$000$}13. & \phm{$0$}0.24 \\
37 & 13 12 27.9\phm{$0$} & +42 42 20\phm{$.0$} & 2.4 & $  87$ & $  2.9$ & $ $       &     &       &                        &                       &         \\
38 & 13 12 28.50 & +42 44 28.9 & 2.1 & $  10$ & $  3.6$ & $K$       & 2.6 &  0.23 &  0.0078 & \phm{$000$}29. & \phm{$0$}7.0 \\
42 & 13 12 28.9\phm{$0$} & +42 46 19\phm{$.0$} & 2.3 & $  32$ & $  3.1$ & $ $       &     &       &                        &                       &         \\
43 & 13 12 28.87 & +42 43 04.8 & 2.1 & $  12$ & $  3.7$ & $I$       & 2.5 &  0.23 &   0.021 & \phm{$000$}11. & \phm{$0$}0.070 \\
45 & 13 12 29.20 & +42 44 38.7 & 3.0 & $   7$ & $  2.5$ & $K$       & 2.5 &  0.39 &   0.020 & \phm{$000$}20. & \phm{$0$}3.5 \\
46 & 13 12 29.69 & +42 43 00.1 & 2.3 & $  11$ & $  3.2$ & $I$       & 1.1 &  0.63 &  0.0072 & \phm{$000$}88. & \phm{$0$}2.0 \\
48 & 13 12 29.9\phm{$0$} & +42 45 14\phm{$.0$} & 2.3 & $   9$ & $  3.1$ & $ $       &     &       &                        &                       &         \\
52 & 13 12 31.76 & +42 42 42.7 & 2.3 & $  24$ & $  3.2$ & $I$       & 3.4 &  0.14 &   0.066 & \phm{$0000$}2.2 & \phm{$0$}0.34 \\
53 & 13 12 31.5\phm{$0$} & +42 42 36\phm{$.0$} & 2.9 & $  27$ & $  2.5$ & $ $       &     &       &                        &                       &         \\
56 & 13 12 31.86 & +42 45 04.1 & 2.4 & $   8$ & $  2.9$ & $K$       & 1.4 &  0.55 &  0.0080 & \phm{$000$}68. &         \\
58 & 13 12 32.78 & +42 43 12.4 & 2.6 & $  15$ & $  2.7$ & $I$       & 1.6 &  0.53 &   0.018 & \phm{$000$}29. & 16. \\
60 & 13 12 34.31 & +42 46 00.3 & 3.3 & $  26$ & $  2.3$ & $I$       & 2.4 &  0.47 &   0.012 & \phm{$000$}38. & \phm{$0$}5.9 \\
63 & 13 12 34.81 & +42 45 05.1 & 2.1 & $  18$ & $  3.7$ & $K$       & 2.6 &  0.21 &   0.017 & \phm{$000$}12. &         \\
\sidehead{Negative Sample}
N\phm{0}0 & 13 12 17.9\phm{$0$} & +42 45 11\phm{$.0$} & 2.2 & $- 17$ & $- 3.3$ & $ $       &     &       &                        &                       &         \\
N\phm{0}1 & 13 12 19.5\phm{$0$} & +42 45 47\phm{$.0$} & 2.2 & $- 13$ & $- 3.4$ & $ $       &     &       &                        &                       &         \\
N\phm{0}2 & 13 12 20.0\phm{$0$} & +42 43 33\phm{$.0$} & 3.3 & $-  6$ & $- 2.3$ & $ $       &     &       &                        &                       &         \\
N\phm{0}3 & 13 12 23.6\phm{$0$} & +42 45 28\phm{$.0$} & 2.1 & $- 10$ & $- 3.4$ & $ $       &     &       &                        &                       &         \\
N\phm{0}4 & 13 12 24.8\phm{$0$} & +42 44 37\phm{$.0$} & 2.6 & $-  7$ & $- 2.6$ & $ $       &     &       &                        &                       &         \\
N\phm{0}5 & 13 12 25.5\phm{$0$} & +42 42 04\phm{$.0$} & 2.2 & $-125$ & $- 3.3$ & $ $       &     &       &                        &                       &         \\
N\phm{0}6 & 13 12 30.5\phm{$0$} & +42 45 37\phm{$.0$} & 1.9 & $- 11$ & $- 4.0$ & $ $       &     &       &                        &                       &         \\
N\phm{0}7 & 13 12 31.46 & +42 44 26.0 & 2.0 & $- 11$ & $- 3.8$ & $I_{814}$ & 2.9 &  0.15 &   0.085 & \phm{$0000$}1.7 & \phm{$0$}1.3 \\
N\phm{0}8 & 13 12 32.9\phm{$0$} & +42 45 13\phm{$.0$} & 2.3 & $-  9$ & $- 3.2$ & $ $       &     &       &                        &                       &         \\
N\phm{0}9 & 13 12 34.23 & +42 43 27.3 & 3.3 & $- 17$ & $- 2.3$ & $I$       & 2.3 &  0.49 &   0.037 & \phm{$000$}13. & \phm{$0$}1.7 \\
N10 & 13 12 35.76 & +42 44 43.7 & 2.0 & $- 26$ & $- 3.9$ & $I$       & 4.2 & 0.032 &  0.0067 & \phm{$0000$}4.7 & \phm{$0$}0.15 \\
N11 & 13 12 38.7\phm{$0$} & +42 44 38\phm{$.0$} & 3.3 & $- 69$ & $- 2.3$ & $ $       &     &       &                        &                       &         \\

\enddata
\tablecomments{
Source identification results
with the probability ratio method described in Sect.~\ref{sect:optid}.
The contents are divided into three sub samples;
the primary, supplementary, and negative samples
(Sect.~\ref{sect:samples}).
Names, coordinates, their one sigma errors ($r_0$),
and total 6.7\,$\mu$m fluxes are extracted from \citet{SKC+03}.
The coordinates of the successfully identified sources
are updated with their optical or near-infrared positions
in the identifying image,
which is indicated in the ID column.
Listed signal-to-noise ratios (S/Ns)
are derived from standard deviations of the co-added images
at the source position using a 7.2\,arcsec aperture.
Thus both photon noise and undersampling effects
are included.
With raster observations,
pixel values could range from $\sim 25$\,\% to $\sim 100$\,\% of source fluxes
depending on exact locations of the sources
on the ISOCAM 6\,arcsec pixels,
giving high standard deviations at source positions.
At 6.7\,$\mu$m,
an Airy disk size is 5.6\,arcsec for the ISO telescope.
Note that errors in total 6.7\,$\mu$m fluxes are much larger 
than that can be inferred from these detection S/N values.
Derivation of total fluxes and their associated errors are described in \citet{SKC+03}.
For each identification,
we list the distance between the 6.7\,$\mu$m source
and the identified source in the ID image ($r_1$).
Two probabilities that the identified source
is a true association ($P_\mathrm{t}$)
or a chance association ($P_\mathrm{c}$),
and their ratio are also shown.
The last column shows the second highest probability ratio
for sources within a distance of 3\,$r_0$
from the original 6.7\,$\mu$m position.
The contents for the negative sample are the same
but with negative flux and S/N values.
}
\end{deluxetable*}	

\def\arraystretch{1.5}			
\tabletypesize{\scriptsize}		
\begin{deluxetable*}{rrlrrrrlrcrrlrcr}	
\tablecolumns{16}
\tablewidth{0pt}
\tablecaption{
Source Identifications at the X-ray, Submillimeter, and Radio Wavelengths
  \label{tab:xsr}
  }
\tablehead{
\multicolumn{2}{c}{ISOCAM}		& &
\multicolumn{4}{c}{\textit{Chandra}}	& &
\multicolumn{4}{c}{SCUBA}		& &
\multicolumn{3}{c}{VLA}			\\ \cline{1-2} \cline{4-7} \cline{9-12} \cline{14-16}
			  \colhead{Name}	& \colhead{6.7\,$\mu$m}		& &
\colhead{$d_\mathrm{X}$}& \colhead{Name}	& \colhead{2--10\,keV}		& \colhead{0.5--2\,keV}		& &
\colhead{$d_\mathrm{S}$}& \colhead{Name}	& \colhead{450\,$\mu$m}		& \colhead{850\,$\mu$m}		& &
\colhead{$d_\mathrm{R}$}& \colhead{Name}	& \colhead{1.4\,GHz}		\\
			  \colhead{\#}		& \colhead{($\mu$Jy)}		& &
\colhead{(\arcsec)}	& \colhead{\#}		& \colhead{($10^{-15}$cgs)}	& \colhead{($10^{-16}$cgs)}	& &
\colhead{(\arcsec)}	& \colhead{[BCS99]}	& \colhead{(mJy)}		& \colhead{(mJy)}		& &
\colhead{(\arcsec)}	& \colhead{FIRST}	& \colhead{(mJy)}
}
\startdata
14 &  19 & & 0.8     & 21      &     $<2.7$ & \phm{0}7.9 & & \nodata &          \nodata & \nodata & \nodata & & \nodata &          \nodata & \nodata \\
17 &  17 & & 1.1     & 18      & \phm{0}4.2 & 15.\phm{0} & & \nodata &          \nodata & \nodata & \nodata & & \nodata &          \nodata & \nodata \\
20 &  20 & & \nodata & \nodata & \nodata    & \nodata    & & \nodata &          \nodata & \nodata & \nodata & & 0.2     & J131223.6+424517 & 3.0     \\
28 &  27 & & \nodata & \nodata & \nodata    & \nodata    & & 7.7     & J131225.7+424350 & $<25$   & 2.4     & & \nodata &          \nodata & \nodata \\
40 &  14 & & 0.5     & 15      & \phm{0}5.7 &     $<4.2$ & & 4.7     & J131228.0+424458 & $<25$   & 2.3     & & \nodata &          \nodata & \nodata \\
57 &  18 & & \nodata & \nodata & \nodata    & \nodata    & & 1.8     & J131232.1+424430 & $<25$   & 3.8     & & \nodata &          \nodata & \nodata \\
62 &  88 & & 0.3     & 10      & \phm{0}7.8 &     $<3.9$ & & \nodata &          \nodata & \nodata & \nodata & & \nodata &          \nodata & \nodata \\

\enddata
\tablecomments{
Source identifications of the mid-infrared (ISOCAM/LW2) sources in the SSA13 field
with X-ray (\textit{Chandra}), submillimeter (JCMT/SCUBA), and radio (VLA/FIRST) sources
\citep{MCBA00,BCMR01,BCS+98,BCS99,WBHG97}.
No data are indicated with '\nodata'.
Flux limits are no detections at a level of 3\,$\sigma$.
Listed offsets $d_\mathrm{X}$, $d_\mathrm{S}$, and $d_\mathrm{R}$
are measured from the optical or near-infrared positions
of the 6.7\,$\mu$m sources in Table~\ref{tab:id}.
}
\end{deluxetable*}		

\clearpage
\def\arraystretch{2.1}			
\tabletypesize{\scriptsize}		
\begin{deluxetable*}{rrrrrlllrclr}	
\tablecolumns{12}
\tablewidth{0pt}
\tablecaption{
Properties of the Mid-Infrared Sources
  \label{tab:prop}
  }
\tablehead{
\colhead{Name}		& \colhead{6.7\,$\mu$m}	&
\colhead{$K$}		& \colhead{$I$}		& \colhead{$B$}		&
\colhead{Type}		& 
\colhead{$z_\mathrm{sp}$, $z_\mathrm{ph}$}	& 
\colhead{SED$_1$}	& \colhead{$\chi^2_1$}	&
\colhead{$\log (M_\mathrm{star})$} 		&
\colhead{SED$_2$}	& \colhead{$\chi^2_2$} \\
\colhead{\#}		& \colhead{($\mu$Jy)}	&
\colhead{(mag)}		& \colhead{(mag)}	& \colhead{(mag)}	&
\colhead{}		&
\colhead{}		& 
\colhead{}		& \colhead{}		&
\colhead{($h_{65}^{-2}$\,M$_\odot$)}		&
\colhead{}		& \colhead{}
}
\startdata
 & & & & & & & & & & & \\
\sidehead{Primary Sample}
 3 &  50 & $ 15.8$ & $ 18.7$ & $ 22.1$ & I       & $0.314^{\phm{+0.0}}_{\phm{-0.0}}$ & \nodata & \nodata & $11.16^{+0.04}_{-0.06}$ & E 8 &   4.7 (d4/4) \\
 6 &  66 & $ 16.1$ & $ 19.0$ & $ 22.1$ & I       & $0.313^{\phm{+0.0}}_{\phm{-0.0}}$ & \nodata & \nodata & $11.08^{+0.04}_{-0.09}$ & Sa 10 &   2.2 (d4/4) \\
 9 &  25 & $ 18.7$ & $ 22.6$ & $ 24.7$ & II      & $1.31^{+0.17}_{-0.15}$ & Sb 8 &   0.0 (d4/4) & $11.39^{+0.13}_{-0.17}$ & Sa 5 &   0.6 (d4/4) \\
10 &  21 & $ 18.8$ & $ 22.7$ & $>25.5$ & II      & $0.914^{\phm{+0.0}}_{\phm{-0.0}}$ & \nodata & \nodata & $10.89^{+0.06}_{-0.09}$ & Sa 6 &   8.5 (d3/4) \\
11 &  19 & $ 17.4$ & $ 20.3$ & $ 24.6$ & I       & $0.732^{\phm{+0.0}}_{\phm{-0.0}}$ & \nodata & \nodata & $11.09^{+0.15}_{-0.05}$ & E 3 &   2.4 (d4/4) \\
13 &  24 & $ 17.3$ & $ 20.6$ & $ 24.5$ & I       & $0.479^{\phm{+0.0}}_{\phm{-0.0}}$ & \nodata & \nodata & $10.99^{+0.04}_{-0.08}$ & E 8 &  13.7 (d4/4) \\
14 &  19 & $ 18.9$ & $ 22.0$ & $ 23.6$ & III     & $1.305^{\phm{+0.0}}_{\phm{-0.0}}$ & \nodata & \nodata & $11.05^{+0.13}_{-0.16}$ & Sa 2 &   0.1 (d4/4) \\
15 &  60 & $ 17.4$ & $ 20.1$ & $ 22.3$ & III     & $0.279^{\phm{+0.0}}_{\phm{-0.0}}$ & \nodata & \nodata & $10.27^{+0.09}_{-0.07}$ & Sb 6 &   2.0 (d4/4) \\
16 &  19 & $ 19.6$ & $ 22.3$ & $ 23.9$ & III     & $1.168^{\phm{+0.0}}_{\phm{-0.0}}$ & \nodata & \nodata & $10.96^{+0.13}_{-0.22}$ & Sb 3 &   2.9 (d4/4) \\
17 &  17 & $ 19.9$ & $ 24.5$ & $>25.5$ & II      & $2.12^{+1.64}_{-0.46}$ & Sb 8 &   0.0 (d3/4) & $11.22^{+0.18}_{-0.18}$ & E 1.5 &   1.2 (d3/4) \\
20 &  20 & $ 19.0$ & $ 23.0$ & $>25.5$ & II      & $1.16^{+0.15}_{-0.10}$ & Arp\,220  &   0.3 (d3/4) & $11.28^{+0.14}_{-0.17}$ & E 4 &   1.8 (d3/4) \\
22 & 135 & $ 13.8$ & $<17.0$ & $ 18.6$ & star    & $0.000^{\phm{+0.0}}_{\phm{-0.0}}$ & \nodata & \nodata & \nodata & \nodata & \nodata \\
25 &  39 & $ 16.4$ & $ 19.1$ & $ 22.4$ & I       & $0.24^{+0.03}_{-0.03}$ & Sa 15 &   0.1 (d4/4) & $10.75^{+0.04}_{-0.09}$ & Sa 11 &   4.3 (d4/4) \\
27 &  45 & \nodata & $>23.7$ & $>25.5$ & II      & \nodata$^{\phm{+0.0}}_{\phm{-0.0}}$ & \nodata & \nodata & \nodata & \nodata & \nodata \\
28 &  27 & $ 18.5$ & $ 21.6$ & $ 24.3$ & III     & $1.038^{\phm{+0.0}}_{\phm{-0.0}}$ & \nodata & \nodata & $11.26^{+0.12}_{-0.21}$ & Sb 6 &  11.4 (d5/6) \\
30 &  70 & \nodata & $ 18.2$ & $ 21.4$ & I       & $0.277^{\phm{+0.0}}_{\phm{-0.0}}$ & \nodata & \nodata & $11.11^{+0.11}_{-0.15}$ & E 8 &   0.1 (d3/3) \\
33 &  17 & $ 19.2$ & $ 23.7$ & $>25.5$ & II      & $1.43^{+0.23}_{-0.22}$ & Sb 13 &   0.0 (d3/4) & $11.28^{+0.14}_{-0.20}$ & E 4 &   0.4 (d3/4) \\
35 &  15 & $ 18.8$ & $ 22.1$ & $ 24.6$ & III     & $1.08^{+0.08}_{-0.11}$ & Sa 4 &   0.0 (d4/4) & $10.98^{+0.16}_{-0.19}$ & Sa 4 &   0.0 (d4/4) \\
39 &  27 & \nodata & $ 22.9$ & $ 25.4$ & II      & $2.96^{+0.38}_{-0.28}$ & Sa 2 &   0.0 (d3/3) & $11.67^{+0.13}_{-0.30}$ & Sa 2 &   0.0 (d3/3) \\
40 &  14 & $ 19.9$ & $ 24.0$ & $>25.5$ & II      & $2.70^{+0.85}_{-0.49}$ & Sb 4 &   0.2 (d4/6) & $11.32^{+0.14}_{-0.21}$ & Sa 2 &   4.4 (d4/6) \\
41 &  46 & $ 16.2$ & $ 18.9$ & $ 22.6$ & I       & $0.489^{\phm{+0.0}}_{\phm{-0.0}}$ & \nodata & \nodata & $11.33^{+0.13}_{-0.09}$ & E 5 &   2.3 (d4/4) \\
44 &  37 & $<17.8$ & $ 20.0$ & $ 22.2$ & III     & $0.38^{+0.25}_{-0.03}$ & Sc 6 &   0.0 (d3/4) & $10.29^{+0.12}_{-0.10}$ & Sc 6 &   0.0 (d3/4) \\
47 &  16 & $ 17.5$ & $ 20.5$ & $ 24.1$ & I       & $0.572^{\phm{+0.0}}_{\phm{-0.0}}$ & \nodata & \nodata & $10.91^{+0.15}_{-0.05}$ & E 4 &   2.9 (d4/4) \\
49 &  15 & $ 19.1$ & $ 22.8$ & $>25.5$ & II      & $1.05^{+0.11}_{-0.17}$ & Arp\,220  &   0.0 (d3/4) & $11.08^{+0.16}_{-0.19}$ & Sa 6 &   1.0 (u1/4) \\
50 &  19 & $ 18.5$ & $ 21.4$ & $ 24.3$ & III     & $0.612^{\phm{+0.0}}_{\phm{-0.0}}$ & \nodata & \nodata & $10.70^{+0.05}_{-0.13}$ & Sb 8 &   0.5 (d4/4) \\
51 &  22 & $ 17.2$ & $ 20.4$ & $ 23.9$ & I       & $0.378^{\phm{+0.0}}_{\phm{-0.0}}$ & \nodata & \nodata & $10.80^{+0.04}_{-0.09}$ & Sa 10 &   9.5 (d4/4) \\
54 & 167 & $ 15.6$ & $ 18.4$ & $ 20.9$ & III     & $0.56^{+0.16}_{-0.26}$ & NGC\,6946  &   0.2 (d4/4) & $11.58^{+0.06}_{-0.07}$ & Sa 4 &   2.7 (d4/4) \\
55 &  12 & $ 18.3$ & $ 20.9$ & $ 22.9$ & III     & $0.839^{\phm{+0.0}}_{\phm{-0.0}}$ & \nodata & \nodata & $10.71^{+0.08}_{-0.10}$ & Sc 4 &   4.2 (d4/4) \\
57 &  18 & $ 19.7$ & $>24.7$ & $>25.5$ & II      & $1.94^{+0.41}_{-0.09}$ & Arp\,220  &   0.0 (d3/6) & $11.03^{+0.37}_{-0.16}$ & E 0.1 &   6.2 (d3/6) \\
59 &  18 & $ 17.2$ & $ 20.1$ & $ 23.6$ & I       & $0.415^{\phm{+0.0}}_{\phm{-0.0}}$ & \nodata & \nodata & $10.88^{+0.04}_{-0.08}$ & E 8 &   0.7 (d4/4) \\
61 &  55 & $ 16.8$ & $ 19.4$ & $ 21.5$ & III     & $0.276^{\phm{+0.0}}_{\phm{-0.0}}$ & \nodata & \nodata & $10.44^{+0.06}_{-0.07}$ & Sa 4 &   6.6 (d4/4) \\
62 &  88 & \nodata & $ 19.3$ & $ 22.4$ & I       & $0.241^{\phm{+0.0}}_{\phm{-0.0}}$ & \nodata & \nodata & $10.64^{+0.04}_{-0.10}$ & Sa 11 &   2.1 (d3/3) \\
64 & 107 & $<13.7$ & $<17.2$ & $ 19.4$ & star    & $0.000^{\phm{+0.0}}_{\phm{-0.0}}$ & \nodata & \nodata & \nodata & \nodata & \nodata \\
\sidehead{Supplementary Sample}
 0 &  19 & $ 16.7$ & $ 19.5$ & $ 22.7$ & I       & $0.314^{\phm{+0.0}}_{\phm{-0.0}}$ & \nodata & \nodata & $10.84^{+0.04}_{-0.09}$ & Sa 10 &   3.6 (d4/4) \\
 1 &  18 & $>19.9$ & $>24.7$ & $>25.5$ & II      & \nodata$^{\phm{+0.0}}_{\phm{-0.0}}$ & \nodata & \nodata & \nodata & \nodata & \nodata \\
 2 &  14 & $>19.9$ & $ 24.4$ & $>25.5$ & II      & $2.23^{+2.78}_{-0.81}$ & Sa 4 &   0.0 (d2/4) & $11.27^{+0.18}_{-0.27}$ & Sa 3 &   0.3 (d2/4) \\
 4 &  18 & $ 19.7$ & $ 24.1$ & $>25.5$ & II      & $2.18^{+1.92}_{-0.52}$ & Sb 6 &   0.0 (d3/4) & $11.39^{+0.18}_{-0.27}$ & Sa 3 &   1.1 (u1/4) \\
 5 &  34 & \nodata & $>23.7$ & $>25.5$ & II      & \nodata$^{\phm{+0.0}}_{\phm{-0.0}}$ & \nodata & \nodata & \nodata & \nodata & \nodata \\
 7 &  66 & \nodata & $>23.7$ & $>25.5$ & II      & \nodata$^{\phm{+0.0}}_{\phm{-0.0}}$ & \nodata & \nodata & \nodata & \nodata & \nodata \\
 8 &  12 & $ 17.9$ & $ 21.7$ & $>25.5$ & I       & $1.028^{\phm{+0.0}}_{\phm{-0.0}}$ & \nodata & \nodata & $10.93^{+0.21}_{-0.20}$ & E 3 &   3.7 (d3/4) \\
12 &  63 & \nodata & $ 22.9$ & $ 24.5$ & II      & $1.93^{+0.35}_{-0.33}$ & Sa 7 &   0.0 (d3/3) & $11.87^{+0.18}_{-0.30}$ & Sa 3 &   2.0 (d3/3) \\
18 &   8 & $ 19.0$ & $ 23.6$ & $>25.5$ & II      & $1.06^{+0.15}_{-0.09}$ & GE  &   0.4 (d3/4) & $10.89^{+0.19}_{-0.34}$ & E 5 &   3.2 (d3/4) \\
19 &  17 & $ 18.1$ & $ 21.7$ & $>25.5$ & I       & $0.76^{+0.08}_{-0.08}$ & Sa 15 &   0.0 (d3/4) & $11.00^{+0.06}_{-0.06}$ & E 5 &   4.2 (d3/4) \\
21 &   9 & $>19.9$ & $ 24.7$ & $>25.5$ & II      & $1.73^{+0.47}_{-0.52}$ & Sa 7 &   0.0 (d2/4) & $11.01^{+0.17}_{-0.28}$ & E 3 &   0.0 (u2/4) \\
23 &   9 & $ 19.5$ & $ 23.4$ & $>25.5$ & II      & $1.08^{+0.18}_{-0.16}$ & Sb 11 &   0.0 (d3/4) & $10.93^{+0.17}_{-0.30}$ & E 5 &   0.3 (d3/4) \\
24 &  12 & $>19.9$ & $ 25.0$ & $>25.5$ & II      & $2.70^{+3.21}_{-1.03}$ & Sb 6 &   0.0 (d2/4) & $11.18^{+0.17}_{-0.22}$ & E 1.5 &   0.1 (u2/4) \\
26 &  10 & $ 19.2$ & $ 23.0$ & $>25.5$ & II      & $1.02^{+0.14}_{-0.18}$ & Sb 11 &   0.0 (d3/4) & $10.97^{+0.18}_{-0.25}$ & E 5 &   0.3 (d3/4) \\
29 &   8 & $>19.9$ & $ 22.6$ & $ 24.2$ & III     & $0.620^{\phm{+0.0}}_{\phm{-0.0}}$ & \nodata & \nodata & $\phm{1} 9.26^{+0.20}_{-0.18}$ & Sb 1 &   0.7 (d3/4) \\
31 &   6 & $>19.9$ & $ 22.7$ & $ 25.6$ & III     & $0.76^{+0.17}_{-0.49}$ & Sc 12 &   0.1 (u1/4) & $10.28^{+0.09}_{-0.25}$ & Sb 6 &   0.3 (u1/4) \\
32 &   9 & $ 19.8$ & $>23.7$ & $>25.5$ & II      & $1.50^{+0.86}_{-0.67}$ & GE  &   0.0 (d2/4) & $10.97^{+0.20}_{-0.24}$ & Sa 4 &   0.0 (u2/4) \\
34 &  45 & \nodata & $ 23.1$ & $ 24.5$ & II      & $2.09^{+0.35}_{-0.49}$ & Sb 6 &   0.0 (d3/3) & $11.76^{+0.18}_{-0.27}$ & Sa 3 &   0.6 (d3/3) \\
36 &   8 & $ 19.4$ & $ 24.8$ & $>25.5$ & II      & $1.39^{+0.21}_{-0.24}$ & GE  &   0.9 (d3/4) & $10.96^{+0.19}_{-0.31}$ & E 4 &   4.7 (d3/4) \\
37 &  87 & \nodata & $>23.7$ & $>25.5$ & II      & \nodata$^{\phm{+0.0}}_{\phm{-0.0}}$ & \nodata & \nodata & \nodata & \nodata & \nodata \\
38 &  10 & $ 17.6$ & $ 21.6$ & $>25.5$ & I       & $0.75^{+0.09}_{-0.08}$ & Sa 15 &   9.9 (d3/4) & $11.17^{+0.05}_{-0.09}$ & E 5 &  16.6 (d3/4) \\
42 &  32 & \nodata & $>23.7$ & $>25.5$ & II      & \nodata$^{\phm{+0.0}}_{\phm{-0.0}}$ & \nodata & \nodata & \nodata & \nodata & \nodata \\
43 &  12 & \nodata & $ 22.7$ & $ 25.5$ & II      & $1.07^{+0.14}_{-0.23}$ & Sa 7 &   0.0 (d3/3) & $10.95^{+0.17}_{-0.23}$ & Sa 5 &   0.5 (d3/3) \\
45 &   7 & $ 18.7$ & $ 22.5$ & $>25.5$ & I       & $0.81^{+0.07}_{-0.10}$ & GE  &   0.9 (d3/4) & $10.85^{+0.06}_{-0.11}$ & Sa 7 &   4.7 (d3/4) \\
46 &  11 & \nodata & $ 23.4$ & $>25.5$ & II      & $3.33^{+1.41}_{-1.19}$ & E 0.8 &   0.0 (d2/3) & $11.25^{+0.21}_{-0.28}$ & E 0.8 &   0.0 (d2/3) \\
48 &   9 & $>19.9$ & $>24.7$ & $>25.5$ & II      & \nodata$^{\phm{+0.0}}_{\phm{-0.0}}$ & \nodata & \nodata & \nodata & \nodata & \nodata \\
52 &  24 & \nodata & $ 23.5$ & $>25.5$ & II      & $1.31^{+0.23}_{-0.34}$ & GE  &   0.0 (d2/3) & $11.48^{+0.18}_{-0.28}$ & E 5 &   0.9 (d2/3) \\
53 &  27 & \nodata & $>23.7$ & $>25.5$ & II      & \nodata$^{\phm{+0.0}}_{\phm{-0.0}}$ & \nodata & \nodata & \nodata & \nodata & \nodata \\
56 &   8 & $ 19.0$ & $ 22.5$ & $ 25.2$ & III     & $0.88^{+0.08}_{-0.17}$ & M\,51  &   0.1 (d4/4) & $10.78^{+0.07}_{-0.10}$ & Sa 6 &   1.3 (d4/4) \\
58 &  15 & \nodata & $ 23.7$ & $>25.5$ & II      & $1.43^{+0.29}_{-0.46}$ & E 5 &   0.0 (d2/3) & $11.22^{+0.19}_{-0.30}$ & E 4 &   0.0 (d2/3) \\
60 &  26 & \nodata & $ 22.2$ & $ 24.0$ & III     & $1.33^{+0.15}_{-1.23}$ & Sc 10 &   0.0 (d3/3) & $11.29^{+0.22}_{-0.32}$ & Sa 3 &   0.1 (d3/3) \\
63 &  18 & $ 18.4$ & $ 23.4$ & $>25.5$ & II      & $1.22^{+0.21}_{-0.11}$ & GE  &   0.3 (d3/4) & $11.32^{+0.17}_{-0.25}$ & E 5 &   5.4 (d3/4) \\

\enddata
\tablecomments{
Properties of the primary and supplementary 6.7\,$\mu$m samples.
Magnitudes at the $K$, $I$, and $B$ bands
are corrected total magnitudes as in \citet{CSHC96}.
Measurements with a 3{\arcsec} aperture
are corrected with an average offset.
No data are indicated with '\nodata'.
Upper flux limits are no detections at a level of 3\,$\sigma$.
Lower flux limits are real measurements for saturated sources
or for sources at the image boundaries.
Except two stars, the remaining sources
are divided into three color types
(Sect.~\ref{sect:color})
using the GRASIL SED library
(Appendix~\ref{sect:GRASIL}).
Spectroscopic redshifts $z_\mathrm{sp}$
\citep{SCHG94,CSHC96,BCMR01}
are listed with no error information,
while photometric redshifts $z_\mathrm{ph}$
are shown with their 90\,\% confidence limits.
Photometric redshifts $z_\mathrm{ph}$
are derived with fits to
the GRASIL evolving templates and local ones
(Sect.~\ref{sect:zph}),
while stellar masses $M_\mathrm{star}$ are derived
with fits to the GRASIL evolving templates only
(Sect.~\ref{sect:SM6.7}).
The results of these two fits
are shown separately
as (SED$_1$, $\chi^2_1$) and (SED$_2$, $\chi^2_2$).
The name of best-fit SED is listed with its age in Gyr
for the case of the evolving SED.
The local template for giant ellipticals is marked as GE.
Just after the $\chi^2$ value,
ancillary information on the fit are shown in parenthesis.
For example, d3/4 means that
3 detections among 4 photometric data
were used to determine the best-fit
and the resulting $\chi^2$ value.
The remaining one flux limit
was checked its consistency to the fit.
In case of u1/4,
one upper flux limit (non-detection)
found to be inconsistent with
the initial best-fit derived with the remaining three detections.
Then, the normalization of the initial fit
was decreased down to the upper limit
and the $\chi^2$ value was evaluated with this re-normalized spectrum.
}
\end{deluxetable*}	

\def\arraystretch{2.}		
\tabletypesize{\normalsize}	
\begin{deluxetable*}{cccccc}	
\tablecolumns{6}
\tablewidth{0pt}
\tablecaption{
Stellar Mass Functions and Stellar Mass Densities
  \label{tab:SM}
  }
\tablehead{
\colhead{$z$ Range} & \multicolumn{4}{c}{$\log \Phi(M_\mathrm{star},z)$}                                                        & $\log \rho_\mathrm{star}(z)$ \\
\colhead{}          & \multicolumn{4}{c}{$(h_{65}^{3}\,\mathrm{Mpc}^{-3}\,\mathrm{dex}^{-1})$}                                  & ($h_{65}$\,M$_\odot$\,Mpc$^{-3}$) \\
                    \cline{2-5}
\colhead{}          & \multicolumn{4}{c}{$\log M_\mathrm{star}$ range}                                                          & \\
\colhead{}          & \multicolumn{4}{c}{($h_{65}^{-2}$\,M$_\odot$)}                                                            & \\
\colhead{}          & \colhead{10.15 -- 10.55} & \colhead{10.55 -- 10.95} & \colhead{10.95 -- 11.35} & \colhead{11.35 -- 11.75} & \colhead{}
}
\startdata
\sidehead{Combined Sample}
0.2--0.5            & $-2.3^{+0.5}_{-0.5}$ (3) & $-2.2^{+0.3}_{-0.4}$ (4) & $-2.3^{+0.3}_{-0.3}$ (5) & \nodata                  & $8.7^{+0.2}_{-0.2}$ (12) \\
0.5--1.2            & \nodata                  & $-2.5^{+0.5}_{-0.6}$ (4) & $-2.9^{+0.3}_{-0.3}$ (6) & $-3.8^{+0.7}_{-1.0}$ (1) & $8.2^{+0.3}_{-0.3}$ (11) \\
1.2--3.0            & \nodata                  & \nodata                  & $-3.3^{+0.4}_{-0.5}$ (5) & $-4.0^{+0.5}_{-0.7}$ (2) & $7.6^{+0.3}_{-0.4}$ ( 7) \\
\sidehead{Spectroscopic Sample}
0.2--0.5            & $-2.5^{+0.4}_{-0.6}$ (2) & $-2.4^{+0.3}_{-0.5}$ (3) & $-2.3^{+0.3}_{-0.3}$ (5) & \nodata                  & $8.6^{+0.2}_{-0.3}$ (10) \\
0.5--1.2            & \nodata                  & $-2.5^{+0.5}_{-0.6}$ (4) & $-3.2^{+0.4}_{-0.5}$ (3) & \nodata                  & $8.1^{+0.4}_{-0.4}$ ( 7) \\
1.2--3.0            & \nodata                  & \nodata                  & $-3.7^{+0.9}_{-1.4}$ (1) & \nodata                  & $6.9^{+0.9}_{-1.5}$ ( 1) \\
\enddata
\tablecomments{
Epoch dependent stellar mass functions $\Phi(M_\mathrm{star},z)$
and stellar mass densities $\rho_\mathrm{star}(z)$.
They are derived from the 6.7\,$\mu$m primary sample.
Two sets of estimates are obtained;
one with the combined (photometric and spectroscopic) sample
and the other with the spectroscopic sample.
One sigma errors are estimated from
uncertainties in $V_\mathrm{max}$ and Poisson statistics.
For stellar mass densities,
errors in stellar mass are also taken into account.
Numbers in parentheses are the numbers of 6.7\,$\mu$m galaxies
used in the calculation.
}
\end{deluxetable*}	

\end{document}